\documentclass[journal]{IEEEtran}

\IEEEoverridecommandlockouts

\usepackage[table]{xcolor}

\usepackage[ruled,vlined]{algorithm2e}
\usepackage{svg}
\usepackage{graphicx}
\usepackage{epstopdf}
\usepackage{amsmath,amsxtra, amsfonts, amssymb}
\usepackage{setspace}
\usepackage{times}
\usepackage[export]{adjustbox}
\usepackage{epsfig}
\usepackage{subcaption}
\usepackage{latexsym}
\usepackage{cite}
\usepackage{url}
\usepackage{tikz,pgf,pgfplots}
\usepackage{multirow}
\usepackage{epstopdf}
\usepackage{breqn}
\usepackage{enumitem}
\usepackage{filecontents}
\usepgflibrary{patterns} 
\usepgflibrary[patterns] 
\usetikzlibrary{patterns} 
\usetikzlibrary[patterns] 

\usepackage{acronym}
\usepackage{comment}
\usetikzlibrary{automata,positioning}

\acrodef{AWGN}{additive white Gaussian noise}
\acrodef{ASM}{advance sleep mode}
\acrodef{bpcu}{bits per channel use}
\acrodef{BS}{base station}
\acrodef{CDF}{cumulative distribution function}
\acrodef{DL}{downlink}
\acrodef{DTX}{discontinuous transmission}
\acrodef{EC}{energy consumption}
\acrodef{EE}{energy efficiency}
\acrodef{FDD}{frequency division duplexing}
\acrodef{FM}{fast mode}
\acrodef{HetNet}{heterogeneous network}
\acrodef{ICT}{information and communication technology}
\acrodef{i.i.d.}{independent and identically distributed}
\acrodef{PDF}{probability density function}
\acrodef{MCS}{modulation and coding scheme}
\acrodef{MDP}{Markov decision process}
\acrodef{ML}{machine learning}
\acrodef{OPEX}{operational expenditure}
\acrodef{PA}{power amplifier}
\acrodef{PAR}{peak-to-average ratio}
\acrodef{QoS}{quality-of-service}
\acrodef{QoE}{quality-of-experience}
\acrodef{RV}{random variable}
\acrodef{RB}{resource block}
\acrodef{RE}{resource element}
\acrodef{RL}{reinforcement learning}
\acrodef{SM}{sleep mode}
\acrodef{TD}{temporal difference}
\acrodef{TTI}{transmission time interval}
\acrodef{uDTX}[$\mu$DTX]{micro-scale discontinuous transmission}
\acrodef{UE}{user equipment}
\acrodef{UL}{uplink}
\acrodef{w.r.t.}{with respect to}

\newcommand{\ev}{\boldsymbol{e}}

\newcommand{\wv}{\boldsymbol{w}}

\newcommand{\ltot}{l_i} 
\newcommand{\lc}{l_i^{({\rm c})}} 
\newcommand{\lmaxc}{l_{\rm{max}}^{({\rm c})}} 
\newcommand{\lb}{l_i^{({\rm b})}} 
\newcommand{\lmaxb}{l_{\rm{max}}^{({\rm b})}} 
\newcommand{\pc}{p_i^{({\rm c})}} 
\newcommand{\pb}{p_i^{({\rm b})}} 
\newcommand{\si}{s_i} 
\newcommand{\ei}{\ev_i} 
\newcommand{\di}{d_i} 
\newcommand{\pt}{p_i} 
\newcommand{\ri}{r_i} 
\newcommand{\ai}{a_i} 
\newcommand{\ith}{i^{\rm th}} 
\newcommand{\Ts}{T_{\rm s}}

\newcommand{\rp}{r_{i,p}} 
\newcommand{\rtp}{\tilde{r}_{i,p}} 
\newcommand{\rd}{r_{i,d}} 

\newcommand{\rjd}{r_{j,d}} 
\newcommand{\rjp}{r_{j,p}} 

\newcommand{\qed}{\nobreak \ifvmode \relax \else
      \ifdim\lastskip<1.5em \hskip-\lastskip
      \hskip1.5em plus0em minus0.5em \fi \nobreak
      \vrule height0.75em width0.5em depth0.25em\fi}

\pgfplotsset{compat=1.16}

\begin{document}

%
\title{Digital Twin Assisted Risk-Aware Sleep Mode Management Using Deep Q-Networks}

%
%

\author{Meysam~Masoudi,
        Ebrahim~Soroush, Jens~Zander, and ~Cicek~Cavdar
\thanks{Copyright (c) 2015 IEEE. Personal use of this material is permitted. However, permission to use this material for any other purposes must be obtained from the IEEE by sending a request to pubs-permissions@ieee.org.}

\thanks{M. Masoudi, J. Zander, and C. Cavdar are with the School
of Electrical and Computer Science, KTH Royal Institute of Technology, Sweden,
 e-mail: \{masoudi,cavdar,jenz\}@kth.se. Ebrahim Soroush is with Zi-Tel Company, e-mail: ebrahim.soroush@gmail.com. This work is supported by European Celtic-Next AI4Green project.}
}

\maketitle

\begin{abstract}
Base stations (BSs) are the most energy-consuming segment of mobile networks. To reduce BS energy consumption, different components of BSs can sleep when BS is not active. According to the activation/deactivation time of the BS components,  multiple sleep modes (SMs) are defined in the literature. In this study, we model the problem of BS energy saving utilizing multiple sleep modes as a sequential \ac{MDP} and propose an online \emph{traffic-aware} deep reinforcement learning approach to maximize the long-term energy saving.  However, there is a risk that BS is not sleeping at the right time and incurs large delays to the users. To tackle this issue, we propose to use a digital twin model to encapsulate the dynamics underlying the investigated system and estimate the risk of decision-making (RDM) in advance. We define a novel metric to quantify RDM and predict the performance degradation. The RDM calculated by DT is compared with a tolerable threshold set by the mobile operator. Based on this comparison, BS can decide to deactivate the SMs, re-train when needed to avoid taking high risks, or activate the SMs to benefit from energy savings. For deep reinforcement learning, we use long-short term memory (LSTM), to take into account the long and short-term dependencies in input traffic, and approximate the Q-function. We train the LSTM network using the experience replay method over a real traffic data set obtained from an operator’s BS in Stockholm. The data set contains data rate information with very coarse-grained time granularity. Thus, we propose a scheme to generate a new data set using the real network data set which 1) has finer-grained time granularity and 2) considers the bursty behavior of traffic data. Simulation results show that using proposed methods, considerable energy saving is obtained, compared to the baselines at cost of negligible number of delayed users. Moreover, the proposed digital twin model can predict the performance of the DQN proactively in terms of RDM hence preventing the performance degradation in the network in anomalous situations.
 \end{abstract}

%


\section{Introduction}

Mobile network operators anticipate a surge in their network energy consumption due to  exponentially increasing network traffic and emerging  new services. Therefore, in order to maintain the network sustainability, \emph{energy efficiency} becomes a  key factor in the future deployment designs \cite{Masoudi2019Green}. Studies show that base stations (BSs) consume around $80\%$ of the overall network energy consumption. Thus, reducing BSs' energy consumption can boost the energy efficiency of the network \cite{ETSI,masoudi2020device,201505_VTC_debaillie,masoudi2022energy}. In particular, base station  sleeping  is one of the
promising ways of reducing energy consumption in the mobile networks. During a day, there are multitudes but short durations in which  no user is connected to the BS while it is active and consumes energy. In such idle durations, BS can put some of its components into sleep and hence save energy, however, it may lead to  longer serving delay for a number of new arrivals. In this study, serving delay is defined as the time that a user is waiting to be served after its arrival. We look into this energy saving opportunity while keeping under control incurred serving delay to the users.

In the literature, the concept of BS sleeping for the purpose of energy saving has been thoroughly investigated \cite{chang2016joint,wu2013traffic,wu2015base,niu2015characterizing,son2011base,kim2017traffic,guo2016delay,chang2018optimal}. The authors in \cite{chang2016joint}, propose a traffic-aware BS sleeping strategy and derive a sleeping policy to maximize the energy saving. In \cite{wu2013traffic}, the authors propose a scheme to put BS into \ac{SM} until a time when a certain number of users are queued in the BS. In a similar setup, the study in \cite{wu2015base} investigates the impact of bursty arrivals on the incurred delay and energy saving. The studies in \cite{ niu2015characterizing,son2011base,kim2017traffic,guo2016delay} investigate the tradeoffs between total energy consumption and overall incurred delay to the users. The authors in \cite{niu2015characterizing}, derive a closed-form energy–delay tradeoff which can be used in designing BS sleeping policies. In \cite{son2011base}, the authors propose a framework for BS energy saving considering user association. In \cite{kim2017traffic}, the authors solve the problem of joint clustering and BS sleeping. In \cite{guo2016delay}, the authors derive an optimal sleeping policy where the policy takes into account  the maximum tolerable delay for the user. They propose a hysteresis-based sleeping strategy and show that there is a linear relation between amount of incurred delay and energy saving  when  the delay constraint is lower than a threshold.   All the aforementioned studies consider only a single level \ac{SM}. 

Recently, multi level \acp{SM} have been introduced in the literature for potentially better energy saving performance at the BSs \cite{andersson2016energy}. According to the (de)activation transition time of BSs' transceiver chain components, the study in \cite{201505_VTC_debaillie} defines 4 levels of \acp{SM} known as \acp{ASM} each with a different (i) time duration, and (ii) energy consumption:  
\begin{enumerate}
    \item {$\rm SM1$:} This \ac{SM} is the fastest, and the most shallow, whose duration is comparable to the symbol time $\Ts$. Whenever a \ac{BS} does not have any traffic over the entire band of sub-carriers during symbol time per antenna, the {PA} can be switched off to save power. This mode is available and known in current technologies as \ac{uDTX}.
    \item {$\rm SM2$:} A slightly deeper sleep level can be reached by switching off one more set of components with slower (de)activation transition time than that of {$\rm SM1$}, if this longer delay can be afforded by the system. In this mode, the transition is on the \ac{TTI} scale, i.e., 1 msec (a duration of a \emph{sub-frame} constituting 2 \acp{RB} or 14 \acp{RE} assuming \ac{FDD} Frame Structure 1 \cite{2014_lte_book_ChrisCox})
    \item {$\rm SM3$:} This mode has a deeper sleeping level but with a minimum duration of a 10 msec frame (10 sub-frames).
    \item {$\rm SM4$:} This is the deepest \ac{SM} with a minimum duration of 100 frames (1 sec).
\end{enumerate}

Implementing such \acp{SM} can save more energy at the cost of  incurring more  delay to the users that have to wait until the BS is back in its active state. Therefore, novel yet efficient schemes are required  to utilize \acp{ASM} without any adverse impact on the offered \ac{QoS}.

The problem of deciding on the sleeping as well as the level of SMs are complex. Machine learning (ML) techniques have shown promising capabilities to efficiently handle the complexity and tune the parameters in systems with ASMs
 \cite{Pees2110Ana,2017_VTC_fatma_tijani,2018_VTC_fatma_tijani,2019_WCNC_fatma_tijani,el2019distributed,Pees2109,pervaiz2018energy,Masoudi2020Reinforcement}. The authors in \cite{Pees2110Ana} evaluate the energy consumption of 5G and beyond BS with multi level SMs. The authors in  \cite{2017_VTC_fatma_tijani,2018_VTC_fatma_tijani,2019_WCNC_fatma_tijani} investigate 4-level SMs with pre-defined order of (de)activation of SMs. They decide on the number of repetitions for each SM.   In \cite{2017_VTC_fatma_tijani} a heuristic algorithm is proposed to implement the \acp{ASM}. They investigate the impact of the synchronization signaling periodicity on the energy saving performance of \acp{ASM}. In \cite{2018_VTC_fatma_tijani}, the authors propose a dynamic algorithm based on Q-learning for SM selection. In \cite{2019_WCNC_fatma_tijani},  the authors propose a  traffic-aware strategy to build a codebook to map the traffic load to the possible actions using Q-learning algorithm. In these studies, the authors assume a pre-defined order of the decisions which may not result in optimal energy saving. The authors in \cite{el2019distributed} relax the pre-defined order of SMs and propose a traffic-aware reinforcement learning algorithm to select SMs. However, they train their network on a fixed traffic pattern. The authors in \cite{Pees2109} use reinforcement learning to adjust the BSs' configurations, e.g., bandwidth and MIMO parameters, to increase the sleeping time of BS. Since control and signaling can limit the energy saving of \acp{ASM}, the study in \cite{pervaiz2018energy} proposes a control/data plane separation in 5G BSs which allows implementing SMs with longer durations. This separation can also be leveraged in other network settings such as co-coverage scenarios when basic coverage cells may carry all periodic control signaling, leaving capacity-enhancing cells with higher ability to sleep and save energy.  {\color{black} The study in \cite{Masoudi2020Reinforcement}, performs BS sleeping for the capacity cell using tabular reinforcement learning method which is trained over real network data. However, the impact of abnormal or bursty arrivals of users on the BS sleeping energy performance are not investigated. }


ML-based algorithms utilize data extracted from the network to learn the network behavior and to make proper decisions to improve the network performance. However, ML-based techniques performance may deteriorate  due to  1) lack of good training 2) abnormal behavior of data that has not been observed before 3) inaccurate design of the algorithm, etc. When the network is experiencing a state that is not observed before, the performance of ML-based techniques is questionable. Furthermore, the traffic pattern may change during time and re-training may be required. It is necessary to tackle these open issues and answer questions such as “Under what circumstances ML-algorithm performance is not reliable in the cellular network?”, "when is it needed to re-train?" and “if an ML-based algorithm is trained in a BS, can it be used in other BSs?” 
These questions are stemmed from a fact that there is always a risk associated  to the use of ML-based algorithms under different circumstances.  Consequently, there is a crucial need to quantify and measure the potential risk of using these algorithms in the network which is a critical challenge and an open problem, not tackled by existing solutions.

Current systems utilize conservative approaches to deal with the risk of activating any energy saving feature in the network. For instance, most energy saving features are never activated, and few are used only at night. Currently, there is no intelligent risk management approach for energy saving,  deployed in the network. In order to determine the proper time of activation/deactivation, threshold-based approaches are used in which when a certain KPI goes below a certain threshold, energy saving feature is turned off. This approach is reactive in which it only considers instantaneous behavior, without involving any prediction, learning, or proactive approaches. However, if the risks can be predicted in advance, these features can be activated or deactivated dynamically and hence the network can benefit more by using these features. 


One way of predicting the risk of using ML algorithms is to utilize the  paradigm of digital twin (DT) \cite{tao2018digital}. A DT is a virtual representation of a physical entity which can mimic the behavior of the physical entity to simulate, predict, forecast behaviour, and possibly control the entity. This representation can be a data driven or analytical model of the system \cite{ma2019digital}. A typical DT-assisted  system has three main parts: 1) physical entities 2) virtual entities and 3) the interaction interface between them \cite{ma2019digital}. DT can be integrated to the next generation of mobile networks, e.g., 6G, to support the realization of the smart control and  optimization of the network \cite{tao2018digital, lu2020communication}.  DT’s applications are quite wide. It can focus on the end-to-end view of the network providing a higher-level analysis of services or network infrastructure, or can focus on a specific domain like the RAN, e.g., \acp{ASM}. In this study, DT is used to realize risk-aware intelligent  \acp{ASM} decision making at the BSs. In particular, DT is composed of a data driven and analytical representations of a BS using ASMs. Using DT, we can predict the performance of the  \ac{ASM} management algorithm in advance, given the required parameters. Based on the performance evaluation, the system can raise an early alarm to deactivate the \acp{ASM} and check for possible algorithm re-training, provided that a risky situation is identified. Risky situation occurs when data used to train the algorithm becomes invalid or network performance is predicted to be below a tolerable threshold determined by the mobile operator. Hence, assisted by DT,  we can identify the times in which using sleep modes will cause unacceptable performance  degradation, i.e., imposing delay to large number of users.


In this paper,  using a real network data set obtained from a Swedish operator, we propose a deep Q-learning (DQN) algorithm and a DT representation of the system to save energy at BS and avoid unexpected performance degradation of the DQN algorithm at the BS. Preliminary version of this study,  is published in \cite{Masoudi2020Reinforcement}. 
The main contributions of this study are summarized as follows: 
\begin{itemize}
    \item We obtain a data set from an operator in Sweden. We propose a framework to generate fine time granularity data from the real network data considering the bursty behavior of cellular network traffic.
    \item We model the problem of sleep mode management in 5G BSs as a Markov Decision Process.  We propose a DQN algorithm to solve the problem of minimizing the energy consumption of a BS using \acp{ASM} while maintaining the users' QoS. Given the traffic input, the proposed algorithm can dynamically decide on the level and duration of \ac{ASM}.
    \item  We propose a DT model-based on the Markov model that can characterize and predict the performance of the proposed sleep mode management algorithm.
    \item We propose a novel metric to quantify the risk of decision making (RDM) and propose a framework to monitor the network and prevent the BS from taking risky decisions. 

    \item We show that combining the  RDM metric, and the proposed framework for risk monitoring, DT can detect the abnormal behavior of traffic and can retrain the algorithm based on new observed data, if required.
    \item Through analytical and simulation results, we show that when the RDM is high or  there is a miss match between DT and physical network observation, the sleeping features can be deactivated temporarily and the algorithm can be re-trained with new data to avoid the risk of incurring large serving delay to the users. 
 
\end{itemize}

The rest of the paper is organized as follows. In Section \ref{sec:sys}, we explain the system model, adopted power and traffic generation models. In Section \ref{sec:solution}, we explain the proposed deep Q-learning algorithm. In Section \ref{sec:markov}, we present the Markov model  of the sleep mode management algorithms. We evaluate the performance of the proposed algorithm in Section \ref{sec:sim}. We conclude the paper in Section \ref{sec:con}.
\vspace{-0mm}
\section{System Model} \label{sec:sys}
\vspace{-0mm}
\subsection{Network Scenario}
 In this study, we consider a non-stand alone 5G deployment in which there are two BSs, namely capacity BS (CapBS) and coverage BS (CovBS). The CovBS is an LTE BS and provides ubiquitous coverage and cell specific signaling. The CapBS is responsible for providing high data transmission rate, on-demand services, and user-based signaling. Since CovBS is responsible for providing coverage and signaling, it cannot take advantage of deep sleep modes. However, CapBS can go to sleep and wake up when it is needed. In this study, we focus on CapBS which from now on we refer to it as BS. 
We also consider that BS can go to SM1-SM3 when there is no user in the network. Table \ref{tab:power_values} presents the power consumption and minimum duration of each SM for macro-base stations with max power of 49 dBm over 3 sectors with a bandwidth of 20 MHz \cite{201505_VTC_debaillie}. 

In mobile networks, BSs are required to communicate with users with periodic signaling that are transmitted in the first and fifth LTE sub-frames. Hence it is not possible to leverage from long and deep SMs. For this reason, 5G New Radio (5G NR) standards allow operators to adjust their periodicity of synchronization signals to 5, 10, 20, 40, 80 and 160 ms, which makes it possible to leverage from deeper/longer SMs. 5G deployments  support standalone and non-standalone deployment. In standalone deployment, operators deploy a separate 5G core and NR for 5G BSs. In non-standalone deployment, the network uses the existing LTE radio access for signaling and existing evolved packet core for backhauling. This approach allows the operators  to introduce  5G new services  while reusing the existing LTE networks. Moreover, in such networks it is possible to separate control and data plane and utilize 5G BSs for data provisioning and use LTE BSs for signaling.

\begin{table}[!t]
\footnotesize
    \caption{\small Power and Minimum Durations from \cite{imec_online_tool, 201505_VTC_debaillie}}
    \label{tab:power_values}
    \centering
    \small
    \begin{tabular}{|c|c|c|c|c|c|}
       \hline
            Mode & \multicolumn{2}{|c|}{\rm Active} &  \multirow{2}{*}{\rm SM1} & \multirow{2}{*}{\rm SM2} & \multirow{2}{*}{\rm SM3}   
            \\\cline{2-3}
            & 100\% & 0\% & &  &
            \\\hline
Pow. (W)& 702.6    & 114.5 &  76.5 & 8.6 & 6.0  
            \\\hline
Dur. (sec)&    \multicolumn{3}{|c|}{$\Ts$: fast modes (FMs)} & 1 msec & 10 msec
            \\\hline
    \end{tabular}
\vspace{-5mm}
\end{table}

\subsection{Power Model}

The adopted BS power consumption model  follows the one implemented  by IMEC in \cite{imec_online_tool} and reported in \cite{201505_VTC_debaillie}. {\color{black} Equation~\eqref{eq:power1} formulates the operational power consumption of a BS and} Table \ref{tab:power_values} summarizes the power consumption values with the configuration parameters summarized in Table \ref{tab:sys_parameters}. {\color{black} When BS is active, it consumes $P_{idle} + P_{t}$ where the first term is the idle power consumption when no SM is activated, and there is no active user serving, and $P_{t}$ is the dynamic power consumption which is proportional to the BS's load. When BS is in sleep mode, the operational power consumption of BS operating in $\text{SM_i}$ is $P_{\text{SM_i}}$ .   In this study, } the BS can operate in three modes, namely fast mode (FM), SM2, and SM3. FM merges three cases: 1) active with $0$ to $100\%$ load, 2) idle with $0\%$ load, and 3) SM1. In this mode, BS makes a decision and hops instantaneously into any operational mode depending on the arriving traffic without any need for optimization. The remaining two sleep modes incur a transition delay from the time of decision until completion of the action, which causes an additional queuing delay of traffic arriving in the transition period. 

\begin{IEEEeqnarray}{rCl} \label{eq:power1}
    P_c =\begin{cases}
    P_{idle} + P_{t} \quad\textit{If BS is in active mode}\\ \\
    P_{\textit{SMi}}\quad\textit{If BS is in SMi,} \quad i \in \{1,2,3\}\\
    \end{cases}
\end{IEEEeqnarray} 
The BS consumes transient power for switching, between operational modes \cite{yang2017hysteretic}.  Denote the total power consumption of switching in duration of $T$ by $P^{T}_{sw}$, power consumption per switching by $P_{sw}$, and the number of switching in duration of $T$ by $F_m$, then we have 
\begin{IEEEeqnarray}{rCl}
    P^{T}_{sw} = F_{m} P_{sw}
\end{IEEEeqnarray}

\subsection{Traffic Model} \label{sec:data}

\subsubsection{Data Set Description}

Cellular traffic varies hourly depending on the daily activity of users in the area. Understanding the traffic patterns of cellular networks is extremely valuable for  better network resource management. In this study, we use a data set from a Swedish mobile network operator. The data set is an anonymized averaged data rate for both uplink and downlink, collected by the operator from multiple BSs in Stockholm for two months in 2018, with one sample taken every 5 minutes. This coarse-grained data set guarantees the credibility of our daily traffic pattern analysis and modeling. This data set can provide us with long term dependencies of network traffic, however, this information is too coarse to provide information about very short time instances comparable to sleep mode duration which is in order of milliseconds. On the other hand, it is not possible to obtain data in such fine-grained granularity due to the limitations in the deployed equipment at BSs  measuring the  KPIs. Therefore, we should adopt a model to generate traffic data with short term dependencies while maintaining the daily pattern and long term dependencies of traffic data.
\subsubsection{Distribution of traffic} \label{sec:datadist}

The data collected by the operator needs to be preprocessed due to the existence of  the incomplete information in the provided data rates at some time instances. Moreover, the information on the average provided rate cannot be directly used in our study, since we need the demand per user. The preprocessing includes the following steps. First, we eliminate the empty and incomplete information from the data set.  Then, we focus on the aggregated downlink traffic of one of such \acp{BS} in one location in Stockholm, only considering the \ac{FDD} downlink band for simplicity, and without loss of generality. Very similar studies can be performed to simultaneously take into account all different \ac{BS} bands, considering both uplink and downlink. This coarse grained original data set is further used as follows to obtain the required information to generate short term traffic data.

For data preparation, we first collate the average data rates of same time index over all days into one set. Denote the set of data at time index $i$ by $\mathcal{O}_{i}$:
\begin{eqnarray}
\mathcal{O}_{i} = \{o_1, o_2, \ldots, o_{\kappa}\}, \quad \textrm{for} \quad i \in \mathcal{T}
\end{eqnarray}
where $\kappa$ is $60$ in our study since we have data for $60$ days and $\mathcal{T}$ denotes the set of the time index of data during the day. The cardinality of set  $\mathcal{T}$ is $24*60/t_s$ where $t_s$ is sampling time which is $5$ minutes in our study. Then, for each set, i.e., each 5 minutes, we calculate the mean, i.e., $E(\mathcal{O}_{i})$, and variance, i.e., $Var(\mathcal{O}_{i})$, of the original data. Finally, based on derived statistics, we can generate demanded data rate for each time step and each user, using the steps explained in the next subsection. It is worth mentioning that this coarse grained time granularity data  provides no information about the time of arrivals or the time duration between two arrivals which is a key parameter to determine the idle time of the BS. Therefore, an accurate and yet tractable model is required to generate the random arrivals of the users. 

\subsubsection{Bursty Behavior of Traffic Data}\label{sec:burst}
In order to model the user arrivals, various distributions such as 
the Poisson distribution \cite{guo2016delay}, log normal distribution \cite{2019_WCNC_fatma_tijani}, hyper exponential distribution \cite{2018_VTC_fatma_tijani}, and interrupted Poisson process (IPP) \cite{liu2018deepnap} have been used in the literature. Among these models, IPP and hyper exponential distribution can model the bursty behavior of incoming traffic while they are  mathematically tractable \cite{liu2018deepnap}. It  can be shown that the IPP is stochastically equivalent to a hyper exponential H2 renewal process \cite{beyer1996predator}.

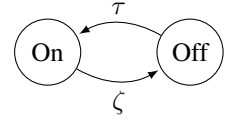
\begin{figure}
    \centering
\begin{tikzpicture}
\node[state]                               (0) {On};
\node[state,right=of 0]                    (2) {Off};

\draw[
    >=latex,
    auto=right,                      
    loop above/.style={out=75,in=105,loop},
    every loop,
    ]
    (0)   edge[bend right]  node {$\zeta$}   (2)
    (2)   edge[bend right] node {$\tau$}   (0)
    
    ;

\end{tikzpicture}

    \caption{\small Data traffic  model as interrupted Poisson process }
    \label{fig:IPP}
\end{figure}

In IPP, as illustrated in Fig. \ref{fig:IPP}, arrivals have two states, i.e., ON and OFF states. In ON state, arrivals follow the Poisson model with parameter $\lambda$ and in OFF state there is no arrival. Assume that the transition rate from ON to OFF (OFF to ON) is denoted by $\zeta$ $\left(\tau\right)$, the steady state probabilities are,
\begin{eqnarray}
P_{\footnotesize\textit{ON}} = \frac{\tau}{\tau+\zeta},\quad
P_{\footnotesize\textit{OFF}} = \frac{\zeta}{\tau+\zeta}
\end{eqnarray}

In order to complete the steps of data generation, during ON state users are generated according to the Poisson distribution with parameter $\lambda$. In OFF state, there will be no arrivals i.e., $\lambda= 0$. Denoting $U$ as the number of arrivals per given time $T$, the mean and variance of $U$ is given by \cite{beyer1996predator}:
\begin{eqnarray} \label{eq:expected_user}
E(U) &=&  \frac{\lambda \tau}{\tau+\zeta} T\\
Var(U)\!\! &=&\!\!  \frac{\lambda \tau T}{\tau+\zeta}  +  \frac{2\lambda^2 \tau \zeta T}{(\tau+\zeta)^3}\left[1 - \frac{\left( 1 - e^{-(\tau+\zeta)T}\right)}{(\tau+\zeta)T}  \right].
\end{eqnarray}
For  $(\tau+\zeta)T\gg1$, the variance can be approximated as, 
\begin{eqnarray} \label{eq:var_user}
Var(U)\!\! &\approx&\!\!  \frac{\lambda \tau T}{\tau+\zeta}\left[1  +  \frac{2\lambda \zeta }{(\tau+\zeta)^2} \right].
\end{eqnarray}

\subsubsection{Data generation} \label{sec:datagen}
In previous two subsections, we derived the statistics of the number of arrivals and the data sets. In this section we explain how to map the parameters of the IPP model so that it matches with the statistics of the original data set.
Denote the data rate request per arrival by $\psi_j$ and the aggregated data rate request per given time $T$ by $\Psi $. 
\begin{eqnarray} \label{eq:psi}
\Psi  &=& \sum_{j=1}^{U} \psi_j\\ \label{eq:expected_psi}
E(\Psi) &=& E(U)E(\psi)\\ \label{eq:var_psi}
Var(\Psi) &=& E(U) Var(\psi) + Var(U) E(\psi)^2.
\end{eqnarray}
In Equation \eqref{eq:psi}, $\psi$-s are independent random variables and for a sufficiently large value of $U$, e.g., more than $30$,  $\Psi$ is normally distributed, regardless of distribution of $\psi$-s, with mean and variance of $E(\Psi)$ and $Var(\Psi)$, respectively. 

In order to generate the user arrivals and their  data rates with the original data set statistics, the model parameters, i.e., $\tau,$ $ \zeta,$ $ \lambda, $ $ E(\psi),$ and $ Var(\psi)$, should be set in a way that Equations \eqref{eq:expected_psi} and  \eqref{eq:var_psi} hold.

Let us assume that $\psi$ is exponentially distributed with mean and variance of $\overline{\psi}$ and $\overline{\psi}^2$, respectively. At each time index $i$, $E(\Psi)$ and $Var(\Psi)$ are set to $E(\mathcal{O}_{i})$, and  $Var(\mathcal{O}_{i})$, respectively. Now, we have 4 unknown parameters and two equations. Since the information of these parameters is not available in the data set, we have to assume two parameters and derive the other two parameters. Let's assume that $\tau$ and $\zeta$ are given \footnote{In the operators' deep packet inspection level data set, the time of service requests and the service durations are available. Hence, the statistics of the  $\tau$ and $\zeta$ can be derived from the data set. }, then by inserting \eqref{eq:expected_user} and \eqref{eq:var_user} into \eqref{eq:expected_psi} and \eqref{eq:var_psi}, $\lambda$ and $E(\psi)$ are derived as follows:  
\begin{eqnarray} \label{eq:lambda}
\lambda &=& \frac{\tau + \zeta}{ \tau T \left[ \frac{Var(\Psi)}{E(\Psi)^2} - \frac{2 \zeta}{\tau T (\tau+\zeta)} \right]}\end{eqnarray}
\begin{eqnarray}
\label{eq:exp}
E(\psi) &=&  \frac{ (\tau  + \zeta) E(\Psi)}{\tau \lambda T}.
\end{eqnarray}

With \eqref{eq:lambda} and \eqref{eq:exp} we have all the parameters to generate the user arrivals with their data rate demand.

\subsection{BS Model}
In this study, we assume that the BS  has two operating states namely, active mode and sleep modes. If BS is active, users are served with rate $\mu$. At the same time, BS can put new service requests with arrival rate of $\lambda$. The BS goes to sleep mode only if the last user in the BS is served. If BS is in sleep mode, all new users must wait to be served until the BS turns into active mode again.

\section{Risk-aware sleep mode management}
\begin{figure}[!t]
    \centering
    \includegraphics[width=.85\columnwidth]{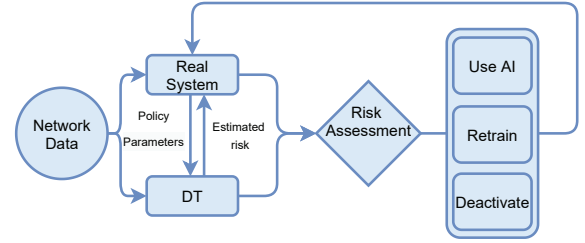}
    \caption{Digital twin assisted decision making}
    \label{fig:abs}
\end{figure}

In this section, we  propose a framework  for  risk-aware sleep mode management.  AI or Machine learning algorithms  are prone to anomalous and unknown data and their performance can be inadequate or sub-optimal. A mechanism is needed to define the risk, monitor the performance of AI with respect to the input traffic to find out whether re-training is needed. As shown in Fig.\ref{fig:abs}, we propose a DT that receives  the network data as an input, communicates with and mimics the behaviour of the real system composed of the physical network and AI module. DT continuously assesses the risk by analyzing and predicting the performance pro-actively. Based on this risk assessment and prediction, the management framework decides whether to use the AI, retrain the network, or temporarily deactivate the  AI module corresponding to deactivating SMs. In the following we explain this approach in detail. First, we introduce the concept of risk associated to the BS sleeping algorithms. We explain the DT model for BS ASM management to estimate the risk. Then,  we combine the intelligent sleep mode management algorithm {\color{black} (SMA)} and DT, and explain the framework for risk-aware BS sleep mode management, as is depicted in Fig.\ref{fig:RDMFlow}. 

\subsection{Risk-Aware Sleep Mode Management} \label{sec:RDM}
In this study, risk refers to a situation in which SM management algorithm takes sleep decisions resulting in delays in connection setup time of future user arrivals. The more users experiencing delay, the higher value of risk should be associated to the situation. Therefore, if the risk could be measured in advance, BS can avoid delaying large number of users. For this purpose, we utilize the  DT concept  as  a virtual representation of the  BS sleeping process.

\begin{figure*}[!t]
     \centering
        \includegraphics[width=0.9\textwidth]{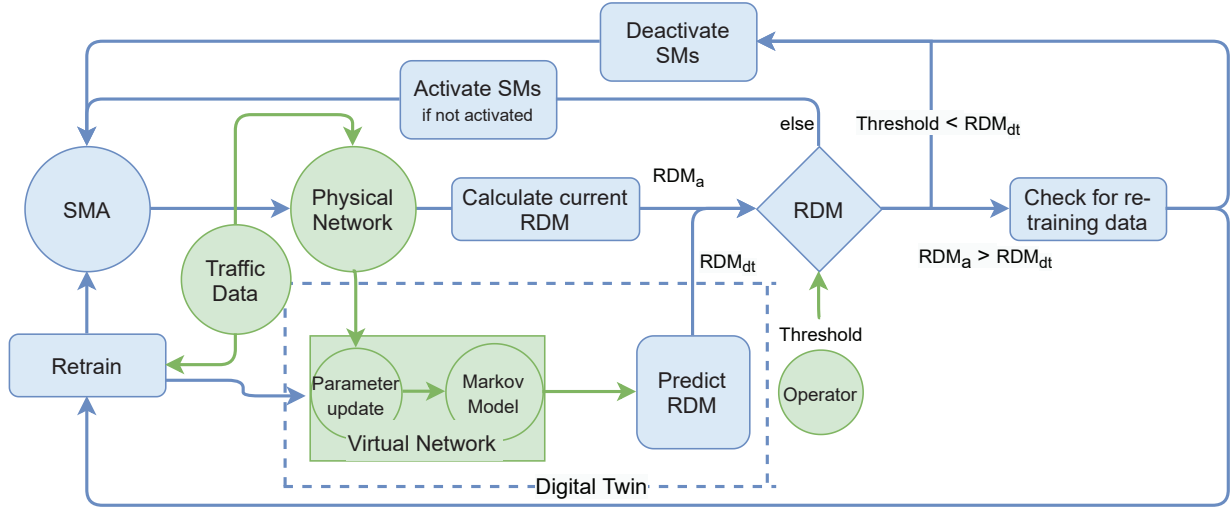}
    \caption{\small Digital twin-based {\color{black} risk-aware} sleep mode management}
    \label{fig:RDMFlow}
\end{figure*}

In Fig.\ref{fig:RDMFlow}, we  illustrate the proposed DT assisted {\color{black}risk-aware} SM management framework. In this scheme, the physical network  provides the required information, e.g., duration of sleeping,  ON/OFF state duration, sleeping time, and arrival rate, to the virtual network to update the model parameters to be used in analytical modeling in DT.  The virtual network is composed of two modules. The first one updates the parameters according to the physical network (only for the initial parameters it uses the training data). The second one is a Markov model with the updated parameters. The virtual network together with the RDM prediction module construct the DT, details of which will be discussed later in Section \ref{sec:markov} and \ref{sec:pm}.   These updated  parameters and the performance metrics are used to calculate the RDM in the network using Equation \eqref{eq:RDM}. The calculated RDM {\color{black} in the DT, i.e., $\text{RDM}_{dt}$, is the expected or predicted value of risk in the next time window,  which is defined as the duration in which the risk is evaluated.  Therefore, the DT can predict the risk of sleeping by utilizing the Markov model and its updated parameters.   Then, the estimated risk is compared with the threshold set by the operator. It will also be compared with the next the actual risk, i.e., $\text{RDM}_a$, which will be available  at the end of the next time window}\footnote{\color{black} The time window should be long enough to collect enough samples from the network to calculate the risk and also short enough to avoid delaying large number of users, e.g., for instance in order of seconds.}. 

{\color{black} If the predicted risk is higher than the predefined threshold,  regardless of  $\text{RDM}_a$, the \acp{SM} should be deactivated. By this mechanism, operator intent is taken into account. Otherwise, if $\text{RDM}_{dt}$ is below the predefined threshold, BS can activate the SMs and benefit from BS sleep mode management algorithms. If the actual risk is higher than the predicted risk,  the algorithm might need retraining. If the actual risk is below the predicted risk, the network decides based on the value of $\text{RDM}_{dt}$. }

This algorithm is designed and explained in Section \ref{sec:solution}. Using the risk monitoring procedure in Fig.\ref{fig:RDMFlow}, the operators  can make sure that their sleep mode management algorithm does not take sleep decisions when there are unexpected user arrivals  or is not missing the opportunity of energy saving when it is safe to save energy.   

\subsection{Digital twin model} \label{sec:dtmo}
In this study, as is depicted in Fig.\ref{fig:RDMFlow}, the digital twin has three main parts, 1) parameter update, 2) network model, and 3) prediction. The former two constructs the virtual network, representative of the physical network. The model is continuously updated from real-time data, and uses machine learning and reasoning to help decision-making. The network is modeled as hidden Markov process and in the prediction phase we use the updated virtual network to estimate and predict the future performance metric of the BS sleeping, e.g., sleeping duration, probability of delaying users, and the risk for each state of the network. In the following we explain the Markov model for BS sleep mode management algorithm. 

\subsection{Virtual Network Model: A Hidden Markov Process } \label{sec:markov}

\begin{figure*}[!t]
    \centering
    \resizebox{.75\textwidth}{!}{

\begin{tikzpicture}

        \tikzset{node style/.style={state, 
                                    minimum width=1cm,
                                    line width=.25mm}}

        \node[node style,fill=gray!60!white] at (0, 0)     (s32)     {$S_{3,2}$};
        \node[node style,fill=gray!60!white] at (3, 0)     (s12)     {$S_{1,2}$};
        \node[node style,fill=gray!60!white] at (1.5, -2)  (s22)     {$S_{2,2}$};
        \node[node style,fill=gray!30!white] at (3, 2)     (s11)     {$S_{1,1}$};
        \node[node style,fill=gray!30!white] at (0, 2)     (s31)     {$S_{3,1}$};
        \node[node style,fill=gray!30!white] at (1.5, -4)     (s21)     {$S_{2,1}$};

        \node[node style,left=of s32]   (A12s3) {$A_{1,2}$};
        \node[node style,left=of s31]   (A11s3) {$A_{1,1}$};
        \node[node style,left=of s22]   (A12s2) {$A_{1,2}$};
        \node[node style,left=of s21]   (A11s2) {$A_{1,1}$};
        \node[node style,right=of s11]   (a11) {$A_{1,1}$};
        \node[node style,right=of s12]   (a12) {$A_{1,2}$};
        \node[node style,right=of a11]   (a21) {$A_{2,1}$};
        \node[node style,right=of a12]   (a22) {$A_{2,2}$};
        \node[right=of a21,text depth=0pt] (h1) {$\ldots$};
        \node[right=of a22,text depth=0pt] (h2) {$\ldots$};
        \node[node style,right=of h1]   (am1) {$A_{M,1}$};
        \node[node style,right=of h2]   (am2) {$A_{M,2}$};
        \draw[every loop,
              auto=right,
              line width=.3mm,
              >=latex]
            (s32)     edge[bend right=30]            node {$\pi_{3,2}$} (s22)
            (s32)     edge[bend right=30, auto=left] node {$\pi_{3,1}$} (s12)
            (s12)     edge[bend right=30]            node {$\pi_{1,3}$} (s32)
            (s12)     edge[bend right=30, auto=left] node {$\pi_{1,2}$} (s22)
            (s22) edge[bend right=30]            node {$\pi_{2,1}$} (s12)
            (s22) edge[bend right=30, auto=left] node {$\pi_{2,3}$} (s32) 
            (am1) edge[bend right=30, auto=right] node {$\zeta$} (am2)
            (am2) edge[bend right=30, auto=right] node {$\tau$}  (am1)
            
            (a11) edge[bend right=30, auto=right] node {$\zeta$} (a12)
            (a12) edge[bend right=30, auto=right] node {$\tau$}  (a11)
            
            (a21) edge[bend right=30, auto=right] node {$\zeta$} (a22)
            (a22) edge[bend right=30, auto=right] node {$\tau$}  (a21)
            
            (a21) edge[bend right=30, auto=right] node {$\zeta$} (a22)
            (a22) edge[bend right=30, auto=right] node {$\tau$}  (a21)
            
            (s11) edge[bend right=30, auto=right] node {$\zeta$} (s12)
            (s12) edge[bend right=30, auto=right] node {$\tau$}  (s11)
            
            (s21) edge[bend right=30, auto=right] node {$\zeta$} (s22)
            (s22) edge[bend right=30, auto=right] node {$\tau$}  (s21)
            
            (s31) edge[bend right=30, auto=right] node {$\zeta$} (s32)
            (s32) edge[bend right=30, auto=right] node {$\tau$}  (s31)
            
            (A11s3) edge[bend right=30, auto=right] node {$\zeta$} (A12s3)
            (A12s3) edge[bend right=30, auto=right] node {$\tau$}  (A11s3)
            
            (A11s2) edge[bend right=30, auto=right] node {$\zeta$} (A12s2)
            (A12s2) edge[bend right=30, auto=right] node {$\tau$}  (A11s2)
            
            (am2) edge[bend left=30, auto=left]             node {$\mu$} (h2)
            (h2) edge[bend left=30, auto=left]             node {$\mu$} (a22)
            (a22) edge[bend left=30, auto=left]          node {$\mu$} (a12)
            (a12) edge[bend left=30, auto=left]          node {$\mu$} (s12)
            (A12s2) edge[bend right=30, auto=right]          node {$\mu$} (s22)
            (A12s3) edge[bend left=30, auto=right]          node {$\mu$} (s32)

            (am1) edge[bend left=30, auto=left]             node {$\mu$} (h1)
            (h1) edge[bend left=30, auto=left]             node {$\mu$} (a21)
            (a21) edge[bend left=30, auto=left]          node {$\mu$} (a11)
            (a11) edge[bend left=30, auto=left]          node {$\mu$} (s11)
            (A11s2) edge[bend right=30, auto=right]          node {$\mu$} (s21)
            (A11s3) edge[bend right=30, auto=right]          node {$\mu$} (s31)

            (h1) edge[bend left=30, auto=left]             node {$\lambda$} (am1)
            (a21) edge[bend left=30, auto=left]             node {$\lambda$} (h1)
            (a11) edge[bend left=30, auto=left]          node {$\lambda$} (a21)
            (s11) edge[bend left=30, auto=left]          node {$\lambda$} (a11)
            (s21) edge[bend right=30, auto=right]          node {$\lambda$} (A11s2)
            
            (s31) edge[bend right=30, auto=right]          node {$\lambda$} (A11s3)
            
            ;
    \end{tikzpicture}}
    \caption{\small Markov model for  advanced sleep mode management.}
        \label{fig:markovChain}
\end{figure*}
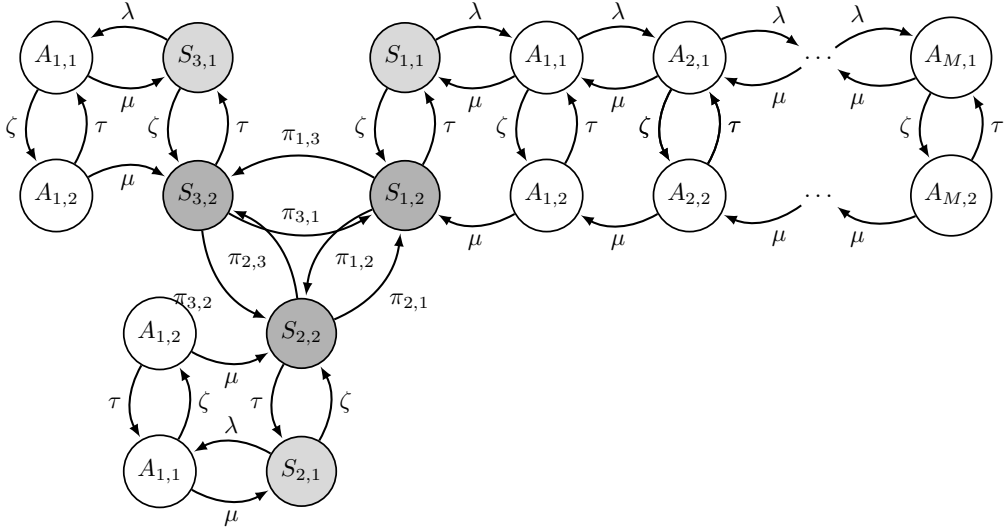

In the DT, we require a realistic analytical model  that can estimate the actual behavior of ML-based BS sleep mode management as a virtual network. In particular, we model the system as a hidden Markov model\footnote{An IPP model plotted in Fig. 1 is a Poisson-emission hidden Markov model with two hidden states \cite{liu2018deepnap} in which one of the states has zero emission rate. Therefore, the state diagram depicted in Fig. \ref{fig:markovChain} is a hidden Markov model. } depicted in Fig. \ref{fig:markovChain} where the states, input traffic, and model parameters are obtained by the interaction with the physical network environment. The environment is defined as the load in the system and level of sleeping. The comprehensive description of the environment is provided in Section \ref{sec:rlbas}. In the following, we present the hidden Markov model, required information, performance metrics, and  a framework for risk monitoring in the system. 

The Markov model should encompass all operating states of the BS to be able to predict the performance of the BS sleep modes. The states contain information on 1) sleep mode level that the BS is in, 2) number of users in active mode, and 3)  whether there are arrivals  or not (ON state or  OFF state as depicted in Fig. \ref{fig:IPP}). We model the state transition diagram of the BS as a Markov model illustrated in Fig. \ref{fig:markovChain}. For instance, state ${S(i,j)}$, $i \in \{1,2,3 \}$ and $j \in \{1,2 \}$, denotes the state that the  BS is in SM$i$ and the traffic arrivals are in the state $j$ where $j=1$ means ON state and $j=2$ means  OFF state. If the BS is in state $A_{m,j}$, $m \in \{ 1,2,\ldots,M\}$ and $j \in \{1,2 \}$, then it is active and $m$ number of users are in the BS. We assume that maximum $M$ number of users can be served and the BS can change its state according to the state transition diagram. When all users in the BS are served, BS goes to one of the SMi-s and stays there until a new user arrives. Before  arrivals of the users, the state changes from $S(i,2)$, OFF state to $S(i,1)$, ON state. When a new user arrives, BS goes from state $S(i,1)$ to state $A(1,1)$ switching from SM$i$ to FM; however, it must wait for the wake-up time of SMi, the duration to activate the sleeping components. When the BS is in any of the sleep modes there is a transition possibility from any of the sleep mode state, i.e., $S(i,j)$ $i \in \{ 1,2,3\}$ to the other SMs if j =2,  there is no traffic arrival.
\subsection{Performance Metrics} \label{sec:pm}
The proposed  Markov model  can enable us to derive the steady state probabilities and later  the calculation of the risk value.  In this section, according to Fig. \ref{fig:markovChain}, given the model parameters, and the power model, we derive the sleeping probability, and we define a new parameter quantifying the risk of taking wrong action.  Let’s define the following probabilities:
\begin{itemize}
    \item $\nu_{i,j}$ be the probability that BS is in state $S(i,j)$.
    \item $u_{m,j}$ be the probability that BS is in state $A(m,j)$.
\end{itemize}
where $S(i,j)$  is the state in which BS is in sleep mode $i$ and the arrival state is $j$ and $A(m,j)$ is the state in which BS is active and there are $m$ number of users being served at the BS and the arrival state is $j$, and $j=1$ means ON state and $j=2$ means  OFF state.

\subsubsection{Probability of Sleeping}
The probability that the BS is in sleep mode is summation of the probabilities that BS is in SM$i$ which is  given by,
\begin{eqnarray} \label{eq:prsleep}
V_s &=& \sum^{2}_{i=1}\sum^{3}_{j=1} \bigg(\nu_{i,j} \bigg)
\end{eqnarray}
In the Appendix \ref{sec:appA}, we have written the balance equations  for Markov model in Fig.\ref{fig:markovChain}. The balance equations are used to calculate the steady state probabilities, i.e., $\nu_{i,j}$ and $u_{m,j}$.

\subsubsection{Number of Switching} 
 Each time BS switches from one sleeping state to another, it adds extra energy consumption or cost to the BS. Therefore, it is crucial to keep track of  the number of switchings which is denoted by $F_m$ and is defined as the number of times that BS switches between SMs or activate/deactivates its components. $F_m$ is derived as, 
\begin{IEEEeqnarray}{lCl} \label{eq:fm}
F_m= 2\mu\sum\limits_{i=1}^{2} u_{1,i} + \sum\limits_{j=1}^{3} \nu_{j,2} (\sum\limits_{k=1/{j}}^{3} \pi_{k,2}).
\end{IEEEeqnarray}
Each time we pass from active to sleep mode in Fig. \ref{fig:markovChain}, i.e., \textit{go to SMi}, later BS has to take a \textit{go to FM} action. The first term in Equation \eqref{eq:fm} counts the number of switching to SMs and wake-ups.
When BS is sleeping and arrivals are in the OFF state, it can switch between SMs. The second term in Equation \eqref{eq:fm} counts the number of switchings between SMs.


{ \color{black}
\subsection{Risk of Decision Making} \label{sec:RDMa}
}
When BS is in  SM and new users arrive, the new arrivals  must wait until the BS wakes up and goes to the active mode again. Denoted by $U_s$,  the average number of users in the BS when BS is any of SMs  is calculated as,
\begin{IEEEeqnarray}{rCl}\vspace{-7pt}
U_s &=& \lambda \nu_{1,1} +\lambda \nu_{2,1} + \lambda \nu_{3,1}
\end{IEEEeqnarray}
where $\lambda$ is user arrival rate and $\nu_{i,1}$, $i \in \{1, 2,3\}$ are the probabilities of being in SM$i$ while arrivals are in  ON state.  From the operators' perspective, the smallest possible value for $U_s$ is more preferable since minimum number of users experience delay until they are being served. Although this metric can reflect the favor of the operators, it is incapable of measuring the performance of the BS sleeping algorithms. For instance,   when the network is busy, i.e. $\lambda$ is high,  the probability of sleeping will be low. However, one inappropriate sleeping decision may result in a large number of users experiencing delay. On the other hand, in off peak hours, when $\lambda$ is low and $V_s$ is high, we may have very few users experiencing delay. Therefore, the parameter $U_s$ cannot solely reflect the higher risk of incurring high delay at peak hours. To tackle this issue, a novel metric called risk of decision making (RDM) is introduced to measure the risk when we take a non-optimal action by taking into account the $U_s$ and the probability of sleeping (in both ON and OFF state).   \textcolor{black}{Risk of wrong decision} is formulated as, 
\begin{IEEEeqnarray}{rCl} \nonumber
 \textit{RDM} &=& E (\textit{Number of users per time  unit} |\textit{BS be in SMs})\\ \label{eq:RDM}
 &=& \frac{ U_s}{V_{s}}
 =\frac{ \lambda  (\nu_{1,1}+\nu_{2,1}+\nu_{3,1}) }{V_{s}}.
\end{IEEEeqnarray}
When the network is in peak hours, $U_s$  might be  high and $V_s$ is low, hence there is not much room to save energy. Since the network is crowded, any wrong sleeping decision may result in a large number of users experiencing delay,  and hence RDM is higher as indicated in Equation \eqref{eq:RDM}. In the off peak hours, $U_s$ might be low and $V_s$ is high. Since less number of users are in the network, RDM is lower which is indicated in Equation \eqref{eq:RDM}. RDM value can also be used to determine the right time/moments to deactivate the BS sleeping algorithm, in order to avoid delaying large number of users. There are two scenarios in which RDM will be high. 

 \textbf{High RDM in Busy Hours}:
 At peak hours, when the network is crowded, it is not beneficial to put BS into sleep modes due to the risk of incurring delay. Therefore, the BS sleeping algorithm can be disabled temporarily until the sleeping becomes beneficial again. It is true that the learning algorithm may learn these moments and avoids sleeping in such events, but it is possible that the algorithm  choose the wrong action due to many reasons such as abnormal behavior of traffic, lack of proper training, and change of traffic pattern.

\textbf{High RDM due to Abnormal Behavior of Traffic}:
When the behavior of users or the traffic pattern changes, e.g., arrival rate changes,  and the algorithm is not previously trained for it, the sleeping decision may yield to unnecessary wake-ups or inappropriate sleepings.  It takes some time for the BS to realize the abnormal behavior of the input traffic which  results in unnecessary energy consumption and/or unacceptable incurred delay. 
When any non-optimal action is taken, it takes some time for the algorithm to converge and  update its policy accordingly. Comparing the RDM with a threshold, we can prevent incurring such delays. The threshold value is set by the operator depending on the hours of the day,  assured QoS, etc.
Using DT, we can estimate the value of RDM. If the experienced value of RDM is above the estimated one, it means the current traffic pattern is different than the expected traffic. Therefore, the BS should disable the sleeping features, i.e., SMs, until either the algorithm is trained over new data set or the traffic behavior becomes normal again.

\subsection{Interaction with DT and Physical Network}

After defining the DT,  underlying Markov model, and the performance metrics, we now explain how the parameters of the  hidden Markov model within the DT can be obtained and updated. The parameters of hidden Markov model is function of user arrival patterns, rates, ON state duration, and OFF state duration. Therefore, these parameters can change during the day. In order to obtain the parameters of the model, i.e., $\lambda$, $\tau$, $\zeta$, from the training data set, we can use a well-known backward-forward algorithm, i.e.,  Baum–Welch algorithm \cite{mccallum2004hidden}. The Baum–Welch algorithm finds the maximum likelihood estimate of the parameters of a hidden Markov model given a set of observed input sequences. This algorithm computes the statistics of the input traffic sequence $O = \{o_1, o_2, \ldots, o_T\}$, and then updates the maximum-likelihood estimate of the model parameters, i.e., ON/OFF state duration and arrival rate. The procedure of Baum–Welch algorithm  to extract the IPP parameters is defined in \cite{liu2018deepnap}.

When BS is in the active mode, BS's transmit power $P_t$ is adapted to match the traffic load. Assume that BS is capable of serving $x$ bit/sec and resources are equally shared between all served users, using a fair scheduler \cite{sousa2020survey}.  The user departure rate is $ \mu = {x}/{l}$ where $l$ is the  user's requested file size and $x = B log (1 + h P_t)$,  and $h = (g/N_{0}B)$ where $g$ represents the channel gain, B is the bandwidth, and $N_0$ denotes the noise density \cite{wu2015base}.

Another set of parameters, i.e.,  $\pi_{i,j}$-s $i \in \{1,2,3\}$ and $j \in \{1,2\}$, are the transition rates between SMs while BS is in OFF state. Meanwhile, this transition rates represent the learned SM management policy. The initial transition rates can be calculated during the training phase. However, during the test phase these rates can be updated by keeping the information about  the average number of times BS switches from one SM to another SM in a given time duration.
The policy can be fed back to hidden Markov model to update the predicted BS sleeping performance metrics, e.g., risk of decision making.



\section{Proposed Sleep Mode Management Algorithm }\label{sec:solution}
\subsection{Reinforcement Learning Elements}
\subsubsection{Action set and Environment}\label{sec:rlbas}
The state of the system at time index $i$, is denoted as $\si$. This state is composed of current traffic and encoded information about previous traffic which is fed back by a recurrent neural network module, explained in Section \ref{sec:lstm}. The current traffic is the number of utilized PRBs at time index $i$. Each action has a transition period, and the system is said to be in the last state until the new state is in action. The action set denoted by $\mathcal{A}$  is defined as 
$\mathcal{ A} = \{{\rm{FM,~SM2,~SM3}}\}.$
Here, $\rm{FM}$ and $\rm{SM k}$, with $k \in \{ 2, 3\}$, denote the go-to-fast-mode and go-to-$\textrm{SM}k$ mode, respectively.

The system environment information at time index $i$ is fully captured via the following 5-tuple:
\begin{eqnarray}
\ei = \left(\pb, \lb, \pc, \lc, \ltot \right) \label{eq:long_state}
\end{eqnarray}
where $\pb$ and $\pc$ are the basic and capacity BS power consumption, $\lb$ and $\lc$ represent the basic and capacity BS \emph{accommodated load} in \acp{RB}, while $\ltot$ is the total traffic load that needs to be served at the $\ith$ time index, respectively. That is $
\ltot = \lb + \lc + \di \label{eq:load}
$
where $\di$ represents the number of delayed \acp{RB} at the $\ith$ time index, which is the difference between the arriving traffic and total served traffic. We also define the total power consumption at the $\ith$ time index as $\pt = \pb + \pc.$ The users arrive to the network as is explained in Section \ref{sec:burst}. At each time index, the amount of load in the environment is determined by the number of users that are in network in the same time index. The value of risk in the environment can be calculated based on the current load, service time and other estimated model parameters.
\subsubsection{Power Saving Reward} When the CapBS is operating in \ac{FM}, no serving delay occurs. However, the system may miss the opportunity of saving more energy if deeper SM is possible. We define a normalized energy saving metric as
\begin{IEEEeqnarray}{rCl}
 \rtp=\frac{\left(p_{SM1}-p_i\right)^+}{p_{SM1}-p_{SM3}},   
\end{IEEEeqnarray}
where $(x)^+$ is a operator that takes the value of $x$ if it is positive and is zero otherwise, $p_{i}, i \in \mathcal{A}$, is the power consumption of SMs and $\rp \in [0,1]$.

\subsubsection{Delay Penalty/Reward}
When the system receives a request while it is in one of the deep SMs, i.e., $\di \neq 0$, or in the case when $\di=\lc=0$, the normalized delaying penalty can be calculated as  
\begin{IEEEeqnarray}{rCl}
   \rd=- \frac{\di}{\lmaxc}, 
\end{IEEEeqnarray}
which takes  $0$ when $\di=0$, and $1$ when $\di=\lmaxc$, respectively.    When the system is in \ac{FM} with $\lc \neq 0$, although no extra power saving is attained, there should exist a reward for the avoided delay, i.e., $(\di =0)\&(\lc \neq 0)$, the reward is
\begin{IEEEeqnarray}{rCl}
    \rd = \frac{\lc}{\lmaxc}.
\end{IEEEeqnarray}
\subsubsection{Total Reward } The incurred reward function at the end of the $\ith$ time index due to the action taken at the end of time index $i-1$ can now be plausibly defined  as
\begin{eqnarray} \label{eq:reward}
\ri =  (1-\alpha) \rp + \alpha \rd, \label{eq:rwrd_fn_simple}
\end{eqnarray}
where $\alpha$ is a weight parameter to prioritize power saving or serving delay. When an action is taken, the system will be frozen for the minimum duration of that action. Therefore, the incurred reward is calculated at the end of such duration. Let $w_{\ai}$ be the transition duration of $a_{i-1}$, as defined in Table \ref{tab:power_values}, the total reward at time index $i$ is calculated as 
\begin{eqnarray}
\overline{r}_i = \frac{\sum_{j=i-w_k-1}^{i} \left((1-\alpha)  \rjp + \alpha \rjd\right)}{w_k}. \label{eq:rwrd_fn_general}
\end{eqnarray}
We assume that when the BS is activated, it stays in $\rm FM$ mode for 14 symbol times to avoid ping-pong effect \cite{elayoubi2011optimal}.
Since every switching from one state to another consumes energy \cite{dolfi2017trade}, we add a negative reward to the total reward if change of action happens, i.e., 
\begin{eqnarray}
\overline{r}_i^{*} =\begin{cases}
 \overline{r}_i \quad \textit{if}\quad a_{i-1} = a_{i}\\\\
 \overline{r}_i - \frac{1}{w_k} \quad else
 \end{cases}
\end{eqnarray}
\subsection{From Q-learning to Deep Q-learning } \label{sec:rl}
Reinforcement learning is a class of machine learning solutions which learns from interactions to achieve a certain goal. The learner, which is called the agent, at each time step (for discrete time) interacts with the environment by taking an action and observes the reward of taken action. Based on the received reward, the agent updates a table, known as Q table, to keep track of values of each action for all states. Q-learning is  an RL algorithm that calculate this value using,
\begin{IEEEeqnarray}{lCl} \label{eq:qlearning}
Q^{New}(s_{i},a_{i}) \leftarrow \\ \nonumber Q^{Old}(s_i,a_i) + \eta \big[ R_{i+1} + \gamma \max_{a}Q(s_{i+1},a_{i+1}) - Q(s_i,a_i) \big]
\end{IEEEeqnarray}
where  $R_{i+1}= \overline{r}_{i+1}$ is the instantaneous reward at time index $i$, $(s_i,a_i)$ is current state-action pair, $\eta \in (0,1]$ is learning rate, and $\gamma \in (0,1]$ is discount factor. 

The computational requirements of Q-learning grow exponentially with the number of states and actions. The problem with large state spaces, which is called the curse of dimensionality, stems from  1) the memory needed for large tables, 2) the time and data needed to fill them accurately 3) many encountered states will be new, will have never been seen before. Therefore, it is not feasible to find the optimal and exact value of Q-function. Alternatively, it is possible to approximate the Q-function with limited computational resources. The function approximation is an instance of supervised learning which can be combined with RL methods to deal with large state space sizes. It helps generalizing learnings from a limited set of past states with a successful approximation over a larger state space.
In this study, instead of using Equation \eqref{eq:qlearning}, we approximate the Q-function using a special type of recurrent neural networks i.e. Long Short Term Memory (LSTM) explained in the following section.
\vspace{-2mm}
\subsection{Deep Q-learning with LSTM} \label{sec:lstm}
When the problem has a large state set, it is not possible to visit all states. In this partially observed environment, the RL agent needs to encode the information about the current input and state-action trajectory.
The most common way of doing this is to use recurrent neural networks (RNNs) which are powerful models for processing sequential data such as time series data. In RNNs, there is a feedback loop which makes the output as part of the input for next time step. With this approach, the information about states can  propagate over time. However, conventional RNNs fundamentally cannot learn long-term dependencies between data sequences. In recent years, LSTM architectures have been gaining popularity due to their ability to model long-term dependencies. An LSTM architecture includes input gate, forget gate, cell state, and output gate. Input gate updates the cell state which can save information of the current state. Forget gate decides whether this information should be kept or discarded, and the output gate decides the next state. With these gates, networks are able to dynamically learn the structure of  longer input sequences and effectively associate memories to adapt which parts of sequence to remember and which parts to forget for the task at hand. Interested readers can refer to \cite{hochreiter1997long} for in-detailed explanation of LSTM architecture.

\begin{figure*}[!t]
    \centering
    \begin{subfigure}{.48\textwidth}
     \centering
        \includegraphics[width=0.95\columnwidth]{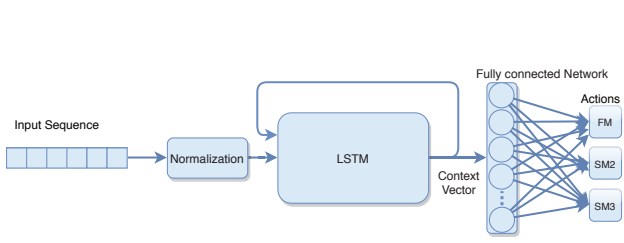}    
        \caption{\small Deep Q-learning non-linear architecture}
         \label{fig:LSTM-a}
    \end{subfigure}
    \begin{subfigure}{.48\textwidth}
     \centering
        \includegraphics[width=0.95\columnwidth]{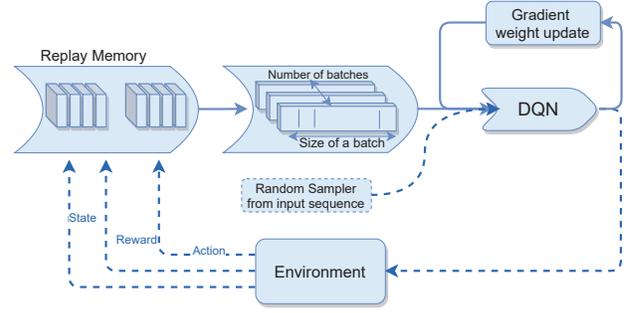}      
        \caption{\small Experience replay training method}
        \label{fig:LSTM-b}
    \end{subfigure}
    \caption{Structure of proposed DQN Sleep Mode Management Algorithm.}
        \label{fig:LSTM}

\end{figure*}

Fig. \ref{fig:LSTM} illustrates the structure of the proposed Deep learning algorithm and the training procedure. According to Fig.  \ref{fig:LSTM-a}, the RL agent gets the normalized input sequence and the context vector, i.e., the mobile traffic data and a hidden vector  (generated in previous time index), as inputs to the first layer of the LSTM module. This module, encodes the current state into a context vector.  This context vector describes the state of the problem in a high dimensional space. The dimension is a hyper-parameter model and the value is chosen depending on  the complexity of the data and problem.  This vector is fed to a fully connected network which connects the context vector to the actions and estimates the value of Q-function for each action. 

To train our DQN model, we use experience replay method \cite{mnih2015human} with two incentives. {\color{black} The first reason is the intrinsic characteristics of sleep mode management problems, such as diverse input traffic patterns. Hence, the proposed method should be online and learn directly from the environment.} Secondly, this method removes the existing correlation in the traffic data sequence, therefore, it makes the training phases more stable \cite{mnih2015human}. As depicted in Fig. \ref{fig:LSTM-b}, in this method, at each time step the agent interacts with the Environment and performs its current actions, then stores the tuple, $(s_i,a_i,R_{i+1},s_{i+1})$,  in the replay memory of length $M$. {\color{black}We randomly select $N$ batches from the replay memory where each batch contains $B$ consecutive samples. Therefore, we have  $N$ uncorrelated batches and a total number of $N\times B$ samples.} We use this batches to train and update the weights of the DQN in the {\color{black} opposite} direction of its gradient. By random sampling, we make sure that the training set is composed of enough {\color{black} span of} samples and our model can generalize well with inexperienced states. 


Since the rewards are normalized, we can use a multi-level binary cross entropy loss function  defined as,    
\begin{IEEEeqnarray}{rCl} \label{eq:loss}
    \mathcal{L} (q,R) = \frac{1}{B}\sum_{b=1}^{B} \ell_{b},
\end{IEEEeqnarray}
 where $q$ is estimated Q-function, $R$ is the reward, B is the batch size, and $\ell_{b}$ is,
\begin{IEEEeqnarray}{rCl}
\ell_{b} = \frac{1}{|\mathcal{A}|}{\color{black}\sum_{a=1}^{|\mathcal{A}|} }\zeta_a \ell_{b,a},
\end{IEEEeqnarray}
where $\zeta_a$ is a rescaling weight given to the loss of each action for loss balancing\footnote{ \color{black} When the number of zeros in data is much higher than the number of other values, the loss due to the more frequent values contributes much more to the total loss. Therefore, the network learns to minimize the loss corresponding to the more frequent values to minimize the loss over the whole data set. However, in our setup, the loss due to non-zero values is more important to be minimized. Therefore, rescaling the loss is needed to avoid this issue.}. {\color{black} We set the $\zeta_a$ to the ratio of non-zero values of the data set to the size of the data set.} $\ell_{b,a}$ is given by,
\begin{IEEEeqnarray}{lCl}\label{eq:lossfunction}
    \!\ell_{b,a} \!=\! -\! \nu_b \left[  R_{a} \!\cdot \log \sigma(q_{b,a})\! +\! (1\! -\! R_{a})\! \cdot \log (1 \!- \!\sigma(q_{b,a}) \right]
\end{IEEEeqnarray}
where $\nu_b$ is a rescaling weight given to the loss of each batch element which is constant in our study.   $\sigma(.)$ is a Sigmoid function to map its input to a value between 0 and 1. $R_a$ and $q_{a,b}$ are rewards due to action $a$ and estimated Q-function for given action $a$ and batch element $b$. In this study, we use Adam optimizer \cite{kingma2014adam} to update the weight parameters to minimize the loss function. Having estimated Q-function in hand, the optimal policy to take an action is 
\begin{IEEEeqnarray}{rCl}
 \pi^{*}(s) = \arg \max \limits_{a}Q(s_{i},a_{i}|\mathbf{v}),   
\end{IEEEeqnarray}
 where $\textbf{v}$ is the context vector calculated by the LSTM layer.

\begin{algorithm}[!t] \small
\KwIn{Traffic data $\mathbf{X}$, batch size $B$,  training data length $L$, prediction length $P_L$, number of epochs $N_P$}
 \textbf{Output:}   For retraining: Training data set, trained DQN model. 
 For SM management: Predicted best action.\\

 \textbf{Initialize:} DQN Model: LSTM layers, Fully connected network\; 
 Normalize traffic data $\mathbf{X}$\;
     \textbf{Train:} \\
    \For{Epoch = $\leftarrow$ 1:$N_P \times N$}{
    \begin{itemize}
        \item To get batch:
            \begin{itemize}
                \item Randomly select $B$ values from $[0,|\mathbf{X}|-(L\!+\!P_L)]$,\\ to form $b_i$-s 
                \item Construct batch data $\mathbf{d_i}$ = $\mathbf{X}\left[b_i:b_i+L-1\right]$ and $\mathbf{f}_i$=$\mathbf{X}[b_i+L:b_i+L+P_L-1]$ to form  $x_i=(\mathbf{d_i},\mathbf{f_i})$ 
            \end{itemize}
        \item Forward batch $x_i$-s to DQN model $M$ \item Get  $a_i$ from DQN
        \item Compute loss based on Eq. \eqref{eq:loss}-\eqref{eq:lossfunction}.
        \item Update DQN model  using Adam optimizer
    \end{itemize}
    }
     \textbf{Test:}\\
         Get \textit{SM active} from Algorithm \ref{alg:RDM};\\
        \While{\textit{SM active}}{
            \begin{itemize}
                \item Get data stream from live network
                \item Estimate Q-function using DQN
                \item Choose best action $a$ using policy: $\pi^{*}(s)=\arg\max \limits_{a}Q(s_{i+1},a_{i+1}|\mathbf{v})$
                \item Perform action $a$
                \item  pass {policy and reward} to BS\;

    \end{itemize}
        }
 \caption{\small DQN sleep mode management algorithm}\label{alg:DQN}
\end{algorithm}

We summarized the procedures of training and testing phases of SMA in the Algorithm \ref{alg:DQN}.
In the Algorithm \ref{alg:DQN}, we explain the training and test phases of the SMA. We assume that the DQN model is given.  For the training phase, we randomly take $B$ values from the data set and feed the data to the DQN model. We get the action as output of the DQN and compute the loss for this action. Then, by using Adam optimizer we minimize the loss and update the DQN model. 

For the test phase of the Algorithm \ref{alg:DQN},  first, the stream of input traffic  is fed to the DQN model and DQN estimates the value of the Q-function for each action. Based on these values, the agent selects the best action and passes the policy to the BS. 

It is worth mentioning that the sleep mode management algorithm is used only when RDM is below the predefined threshold. In other words, sleep mode management algorithm will be in use according to the Algorithm \ref{alg:RDM}. This algorithm explains the procedure of RDM monitoring in the network. According to this algorithm, the network information such as, average arrival rates and transition probabilities, is given to the BS. The BS uses this information and updates the model parameters and predicts the RDM. The value of RDM is used to check if the risk is above or below the threshold. If the value of RDM is above threshold, the SMs are deactivated. If the value of RDM is below the threshold the SMs can be activated. It is important to keep track of the RDM value for a while, let’s say $T_w$, to make sure that the decisions are not made based on instantaneous changes in the network. To avoid spontaneous decisions, moving average operator can be used to cancel out the instantaneous changes in data traffic and RDM. Therefore, the sleep mode management algorithm comes back to loop only if the average value of RDM is low and it is safe to save energy. 
\begin{algorithm}[!t] \small
 \KwIn{Traffic data, transition rates between SMs}
 \SetKwInOut{}{}{}
  \textbf{Output:}    Decide to run which part of Algorithm 1 or not run Algorithm 1.  \\
 Condition=True\;
 \While{Condition}{
     Network status update:
    \begin{itemize}
        \item Update the parameters of hidden Markov model.
        \item Calculate the estimated RDM, i.e., $\text{RDM}_{dt}$.
        \item Calculate the actual RDM in the network, i.e., $\text{RDM}_{a}$\;
    \end{itemize}
  \If{$\text{RDM}_{a}$ or $\text{RDM}_{dt}$ is higher than threshold}{
   Deactivate the SMs\;
   }
   \ElseIf{$\text{RDM}_{a}$ $>$ $\text{RDM}_{dt}$}{
   \begin{itemize}
        \item Deactivate the SMs\;
        \item Run Algorithm 1: Train\;
    \end{itemize}
  }
  \Else{
  \begin{itemize} \small
        \item Monitor $\text{RDM}_{a}$\ for duration $T_w$;
        \item if $\text{RDM}_{a}<$ threshold: Activate the SMs.
        \end{itemize}
  }
  \If{SM active}{
  Run Algorithm \ref{alg:DQN}: Test.}
 }
 \caption{\small Risk management algorithm in the network}
 \label{alg:RDM}
\end{algorithm}
\section{Simulation Results} \label{sec:sim}

\begin{table}[!t] \footnotesize \vspace{0mm}
    \caption{\small Studied System Configuration Parameters}
    \label{tab:sys_parameters}
    \centering
    \begin{tabular}{|c|c|c|} 
    \hline
    \multicolumn{2}{|c|}{Parameter}     & Value 
    \\\hline
    \multicolumn{2}{|c|}{Modulation}    & 16-QAM (4 bits per RE), $M =16$ 
    \\\hline
    \multicolumn{2}{|c|}{Antennas}      & 1 per cell sector, one sector per cell    
    \\\hline
    \multicolumn{2}{|c|}{Bandwidth}     & 20 MHz for CovBS, CapBS
    \\\hline
    \multicolumn{2}{|c|}{{RBs per band}}  & {100 per 20 MHz (LTE-A Compatible)}
    \\\hline
    \multicolumn{2}{|c|}{REs per TTI} & 14 (Based on FDD Frame Structure 1)
    \\\hline
    \multicolumn{2}{|c|}{Symbol Time}  & $\Ts = \frac{1}{14 \times 1000}$ sec $\approx 72 \mu$sec
    \\\hline
    \multicolumn{2}{|c|}{Hidden-dimension/ Batch size} & 256, 200
    \\\hline
    \multicolumn{2}{|c|}{Episode per epoch } & 2048
    \\\hline
    \multicolumn{2}{|c|}{$\tau$, $\zeta $} & 0.1, 0.5
    \\\hline
    \multicolumn{2}{|c|}{\color{black}Risk Threshold } & 1.2
    \\\hline
    \end{tabular}
\end{table}
An event-based numerical simulation is implemented in Python, running the proposed adaptive algorithm and its baseline counterparts. The simulation parameters are defined in Table \ref{tab:sys_parameters}.

\subsubsection{Baselines}
We compare the performance of  the SMA algorithm  with  optimal BS sleeping (OBS) where future traffic arrivals are assumed to be known as an upper bound on energy saving gain {when no delay is incurred to the users}. We follow the procedure explained in Algorithm \ref{algo:optimal_sleep} \cite{Masoudi2020Reinforcement}. 

{\color{black}   In OBS, the BS is fully aware of all future incoming users and their activity times. Scanning all inactivity periods, BS can fill the inactive time with the proper SM.  After the initialization step, OBS fills the inactive time with the deepest possible SM, i.e., SM3. Then, the remaining inactive time will be filled by the next deep SM, i.e., SM2. The remaining inactive time will be filled by SM1. The OBS algorithm, keeps track of the number of actions for each SMs and the energy performance of the BS.  }

This procedure guarantees the highest energy saving gain with no incurred delay to users, yet utilizing the future knowledge{\footnote{\color{black}Since the output of OBS depends on the future inputs and not just past and current inputs, the method is non-causal.}}. {\color{black} We compare the performance of SMA with fixed sleep mode where only $\rm SM1$, i.e., the shallowest SM, is used. SM1 is already implemented in the BSs and can be activated without  requiring any intelligence. In this scheme, BS can go to SM1 when no user is served. } In these baselines all users are served instantaneously with no incurred delay. We also compare the results with Q-learning algorithm  in \cite{el2019distributed}.



\begin{algorithm}[!t]
{\color{black}    
\small
\caption{\small{Optimal \ac{BS} Sleeping}}
\label{algo:optimal_sleep}
\KwIn{\\ Cell load $l^{(c)} = \max\{0,l - \lmaxb\}$ for every symbol time over the entire time horizon. SM counter vector.
Duration of SMs in terms of symbol time, i.e., $\wv=(w_1, w_2, w_3) = (1, 14, 140)$.\\}
 \textbf{Output:}    Energy performance, total SM counter. \\
\textbf{Initialization:} \\
    \begin{itemize}
        \item Set $\wv = (w_1, w_2, w_3) = (1, 14, 140)$.
        \item Set total SM counter, TSMC= $[0,0,0]$.
        \item $\boldsymbol{z}$: A vector containing the run lengths of zero~loads (in symbol times).
        \item  $\boldsymbol{e}$:  A vector with the same length as $\boldsymbol{z}$ containing the energy consumption of the each duration in $\boldsymbol{z}$.
        \item $j=0$,  an index that runs over  entries of $\boldsymbol{z}$.
    \end{itemize}
    \textbf{{SMs} Fitting:}\\

     Set: $\boldsymbol{n}[j]=\boldsymbol{z}[j]$\\       
    \For{ $j=1:1: length(\boldsymbol{z})$}{
    $\text{SMC}= 0$\\
        \For{$k = 3:-1:1$}{
            \While{$\boldsymbol{n}[j] \geq w_k$}{
            $\text{SMC}[k] += 1$,\\
            $\text{TSMC}[k] += 1$,\\

            $\boldsymbol{n}[j] -= w_k$
            }
    }
   
    \textbf{Energy Evaluation :}\\
    Update the energy consumption, $\boldsymbol{e}$ using  Equations (1) and (2).
    }
}    

\end{algorithm}

\subsubsection{LSTM Hyper-parameters}
 {\color{black} Choosing the right hyper-parameters is crucial to design a network, including the deep learning model with the best performance.} Through simulations, we swipe over  different parameters and compare the  performance of the whole network, i.e., loss and accuracy, which is summarized in  Table \ref{tab:lstmpar}.   The loss function is defined in \eqref{eq:loss} and to calculate the accuracy we compare the decision made by the optimizer with the correct decision to make. We pick the parameter values with the best performance  for the rest of simulations. The best performance is achieved with a two-layer LSTM and hidden vector of size $50$ and batch size of $200$.


\begin{table}[!t]
\rowcolors{2}{gray!50}{gray!10}
    \centering
    \footnotesize
        \caption{\color{black} Hyperparameter values of the network}
    \label{tab:lstmpar}
    \begin{tabular}{|c|c|c|c|c|c|c|}
    \hline
    \multicolumn{2}{|c|}{LSTM Performance} &\multicolumn{3}{c|}{LSTM Parameters} \\ \hline
    accuracy& loss  & layers  & hidden size & seq length  \\ \hline 
       0.96940 & 0.08858 & 1 & 50  & 100 \\\hline
                0.96972 & 0.08706 & 1 & 100  & 100 \\\hline
                0.96924 & 0.08882 & 1 & 150  & 100 \\\hline \rowcolor{green!50}
                0.97964 & 0.07855 & 2 & 50  & 100 \\\hline
                0.96000 & 0.09223 & 2 & 100  & 100 \\\hline
                0.96068 & 0.09214 & 2 & 150  & 100 \\\hline
                0.96024 & 0.09438 & 3 & 50  & 100 \\\hline
                0.95708 & 0.09212 & 3 & 100  & 100 \\\hline
                0.96712 & 0.08960 & 3 & 150  & 100 \\\hline
    \end{tabular}

\end{table}


\subsubsection{Data Generation}
\begin{figure*}[!t]
    \begin{subfigure}{.33\textwidth}
        \centering
        \resizebox{1\textwidth}{!}{\input{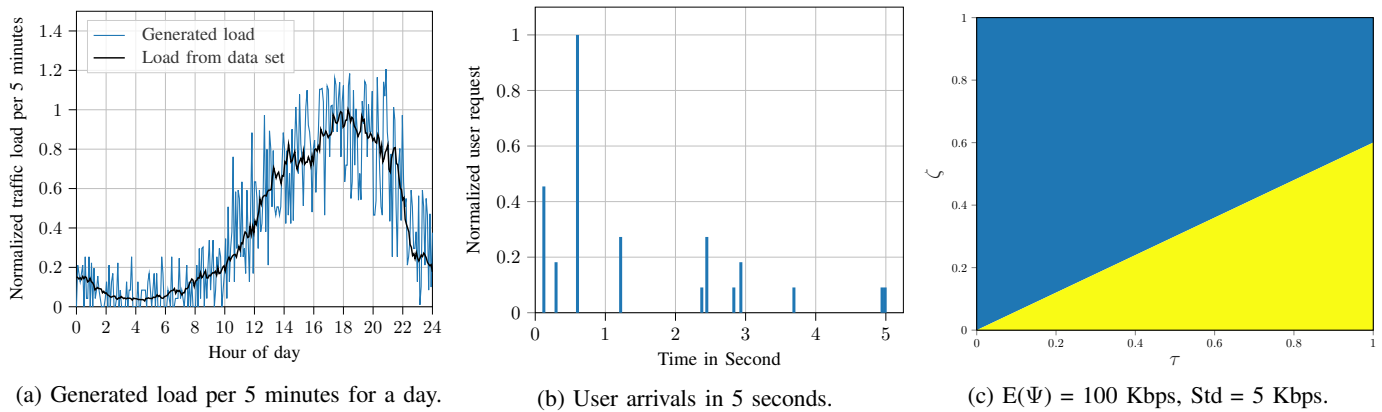}}
        \caption{\small Generated load per 5 minutes for a day. }
    \label{fig:oneminute}
    \end{subfigure}
    \begin{subfigure}{.33\textwidth}
        \centering
        \resizebox{1\textwidth}{!}{\begin{tikzpicture}
\definecolor{color0}{rgb}{0.12156862745098,0.466666666666667,0.705882352941177}
\begin{axis}
[
    every axis plot post/.style={mark options={fill=color0}},   
    xmin=0, xmax=5.25,
    ylabel={Normalized user request},
    ymin=-0.05, ymax=1.05,
    xlabel={Time in Second},
    ymin=0,
    ymax=1.1,
    tick align=outside,
    tick pos=left,
    x grid style={black},
    xtick distance={1},
    ytick distance={.2},
    x grid style={lightgray},
    xmajorgrids,
    xminorgrids,
    xtick style={color=black},
    ymajorgrids,
    yminorgrids,
]
\addplot[ycomb,color0,  ultra thick] table [x={n}, y={xn}] {data.dat};
\end{axis}
\end{tikzpicture}}
        \caption{\small User arrivals in 5 seconds. }
    \label{fig:finegran}
    \end{subfigure}
    \begin{subfigure}{.34\textwidth}
    \resizebox{1\columnwidth}{!}{
%
%
\definecolor{color0}{rgb}{0.12156862745098,0.466666666666667,0.705882352941177}
\definecolor{mycolor1}{rgb}{0.24220,0.15040,0.66030}%
\definecolor{mycolor2}{rgb}{0.97690,0.98390,0.08050}%
\begin{tikzpicture}

\begin{axis}[%
width=4.521in,
height=3.566in,
at={(0.758in,0.481in)},
scale only axis,
xmin=0,
xmax=1,
xlabel style={font=\color{white!15!black}},
tick align=outside,
    tick pos=left,
xlabel={\LARGE $\tau$},
ymin=0,
ymax=1,
ylabel style={font=\color{white!15!black}},
ylabel={\LARGE $\zeta$},
axis background/.style={fill=white},
xtick distance={.2},
ytick distance={.2},
]

\addplot[fill=color0, draw=none, forget plot] table[row sep=crcr] {%
0	0\\
1	0.600266933600267\\
1	1\\
0	1\\
0	0\\
};

\addplot[fill=mycolor2, draw=none, forget plot] table[row sep=crcr] {%
0	0\\
1	0.600266933600267\\
1	0\\
0	0\\
};

\addplot[fill=none, draw=black, forget plot] table[row sep=crcr] {%
0	0\\
1	0\\
1	1\\
0	1\\
0	0\\
};

\end{axis}

\end{tikzpicture}%
}
        \caption{ \small $\text{E(}\Psi\text{) = 100 Kbps, Std = 5 Kbps}$. }
    \label{fig:ipp_param1}
    \end{subfigure}
    \caption{ \small Network load generation as input to the network for $\tau = 0.1$ and $\zeta = 0.5$. Feasible region for $\lambda$ for $E(\Psi) = 100$ Kbps. Yellow shows the feasible and blue represents the infeasible region.}
\end{figure*}

 We use mobile traffic data provided by a Swedish operator. As explained in Section \ref{sec:data}, data sets needs to be processed to be used in our simulation tools.
  In Fig. \ref{fig:oneminute}, we plot the normalized average load  per 5 minutes for original data set and the generated load for the same 5 minutes. The comparison shows that the generated data preserves the average daily behavior of the original data. The fluctuations are  due to the randomness of arrivals and fluctuations in the data set. We can see that the generated load in 5 minutes, which is the finest available time  granularity, is following the trend in the data set. In Fig. \ref{fig:finegran} we show the arrivals in time scale of 5 seconds.

 According to Section \ref{sec:datagen}, the parameters of IPP model, i.e., $\tau,$ $\zeta,$ $\lambda,$ as well as the statistical parameters of distributions of request per arrivals must be set in a way that  mean and variance of the  generated load  matches with the mean and variance of the original data set, for the same duration. However, according to Equation \eqref{eq:lambda}, for a given $\tau$ and $\zeta$, there might be no solution for $\lambda$ and $E(\psi)$. If the ratio of $\frac{var(\Psi)}{E(\Psi)^2}$, i.e., dispersion ratio of data set, is lower, the feasible region for $\lambda$ is wider.  {\color{black} In Fig. \ref{fig:ipp_param1}, we illustrated the feasible region of data generation parameters for the mean value of $100$ Kbps and variance of $5$ Kbps. As can be seen from this figure , when $\tau$  is larger (or $\zeta$ is smaller), i.e., duration of ON state is higher, the feasible region for $\lambda$ is larger. This figure also illustrates that if the combination of $\tau,$ $\zeta,$ and $\lambda,$ lies in the infeasible region, the data generation is not reliable. Moreover, during the test phase if similar situation happens, one cannot rely on the prediction of the parameters $\tau,$ $\zeta,$ $\lambda,$ and the associated risk.}


\subsubsection{Learning performance}
\begin{figure}
    \centering
     \resizebox{.85\columnwidth}{!}{\input{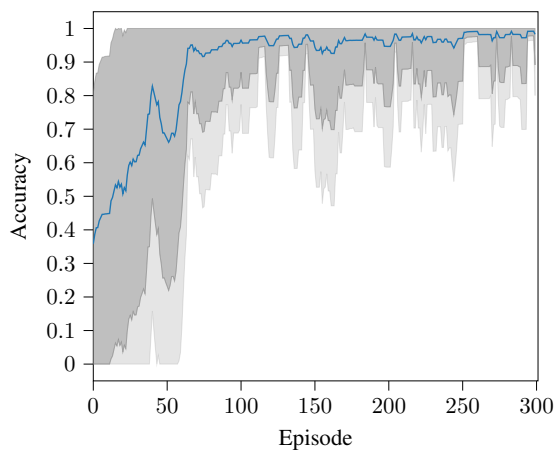}}
    \caption{\color{black}Learning curve the DQN model for alpha = $0.7$. Dark shaded region depicts the one standard deviation and the light shaded are depicts the $3$ standard deviations variations.  }
    \label{fig:conv}
\end{figure}

{\color{black} In Fig.\ref{fig:conv},  we present the learning curve of the proposed DQN algorithm. We compared the DQN selected action with the pre-calculated best action based on the chosen alpha to decide whether the action was correct within an episode. In this figure, each episode contains a sequence of states, actions, and rewards with a sequence length of 100. The solid blue line shows the average ratio of correctly chosen actions for 100 simulation runs of each episode. The dark-shaded and light-shaded areas show the one and three standard deviations of the accuracy for the given episode. A high standard deviation or larger shaded area indicates the uncertainty in the model to select the actions and hence may take different actions at a specific episode. A low standard deviation indicates that the algorithm is more determined and tends to choose similar actions at a particular episode. In the early episodes, the model learns the best actions by exploration and choosing various actions and, hence, high standard deviation. After 100 episodes, the DQN model is adequately trained in this setup.} 

\subsubsection{Energy saving vs incurred delay}
\begin{figure}[!t]
    \centering
    \resizebox{.85\columnwidth}{!}{
\begin{tikzpicture}

\definecolor{color0}{rgb}{0.12156862745098,0.466666666666667,0.705882352941177}
\definecolor{color1}{rgb}{1,0.498039215686275,0.0549019607843137}
\definecolor{color2}{rgb}{0.172549019607843,0.627450980392157,0.172549019607843}
\definecolor{color3}{rgb}{0.83921568627451,0.152941176470588,0.156862745098039}

\begin{axis}[
legend cell align={left},
legend style={
  fill opacity=0.8,
  draw opacity=1,
  text opacity=1,
  at={(0.01,0.5)},
  anchor=west,
  draw=white!80!black
},
tick align=outside,
tick pos=left,
xtick distance={.1},
ytick distance={.1},
x grid style={lightgray},
xlabel={\small Alpha},
xmajorgrids,
xminorgrids,
xmin=-0.05, xmax=1.05,
xtick style={color=black},
y grid style={lightgray},
ylabel={\small Normalized Energy Saving},
ymajorgrids,
yminorgrids,
ymin=0.2, ymax=1.03,
ytick style={color=black}
]
\addplot [semithick, color2, mark=square*, mark size=2, mark options={solid}]
table {%
0 1
0.111111111111111 1
0.222222222222222 1
0.333333333333333 1
0.444444444444444 1
0.555555555555556 1
0.666666666666667 1
0.777777777777778 1
0.888888888888889 1
1 1
};
\addlegendentry{OBS}
\addplot [semithick, color0, mark=*, mark size=2, mark options={solid}]
table {%
0 1
0.111111111111111 0.99
0.222222222222222 0.95
0.333333333333333 0.94
0.444444444444444 0.76
0.555555555555556 0.67
0.666666666666667 0.66
0.777777777777778 0.65
0.888888888888889 0.45
1 0.4
};
\addlegendentry{SMA}
\addplot [semithick, color1, mark=triangle*, mark size=2, mark options={solid,rotate=180}]
table {%
0 1
0.111111111111111 0.99
0.222222222222222 0.9
0.333333333333333 0.89
0.444444444444444 0.52
0.555555555555556 0.5
0.666666666666667 0.47
0.777777777777778 0.4
0.888888888888889 0.4
1 0.4
};
\addlegendentry{Q-learning \cite{el2019distributed}}
\addplot [semithick, color3, mark=diamond*, mark size=2, mark options={solid}]
table {%
0 0.4
0.111111111111111 0.4
0.222222222222222 0.4
0.333333333333333 0.4
0.444444444444444 0.4
0.555555555555556 0.4
0.666666666666667 0.4
0.777777777777778 0.4
0.888888888888889 0.4
1 0.4
};
\addlegendentry{Only SM1}
\end{axis}

\end{tikzpicture}}
    \caption{\small Energy saving as a function of tuning parameter $\alpha$ in the reward function.}
    \label{fig:alphavar}
\end{figure}
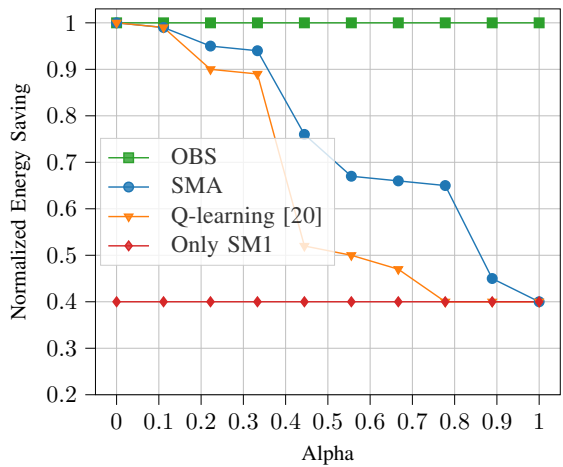

As per Equation \eqref{eq:reward}, the parameter $\alpha$ is a tuning parameter adjusting the weight given to power saving vs minimization of delay, in the reward function.  In Fig. \ref{fig:alphavar}, we present the performance of the SMA with regards to the $\alpha$. The higher  $\alpha$ is, the more weight is put to minimize delay and less weight is assigned to energy saving. Hence, with higher $\alpha$ less energy is saved. We  compare  the performance of SMA with  optimal BS sleeping (OBS) algorithm defined in Algorithm \ref{algo:optimal_sleep}, as an upper bound on energy saving. OBS is always optimal and non-causal due to the knowledge of future information. For $\alpha =0$, where the total reward includes only energy saving reward, SMA can achieve the optimal energy saving but at cost of incurring  delay to the users. When latency is more prioritized, the BS becomes more conservative to choose deeper SMs and prefers shallower SMs. Therefore, less energy is saved in favor of less incurred delay. SMA outperforms the fixed SM where only SM1 is in use and BS cannot benefit from deeper SMs. Moreover, SMA performs better than Q-learning algorithm \cite{el2019distributed} because SMA can find better long-short term dependencies in traffic data and hence leverage this information to find a better energy saving policy.

\subsubsection{Daily energy saving}

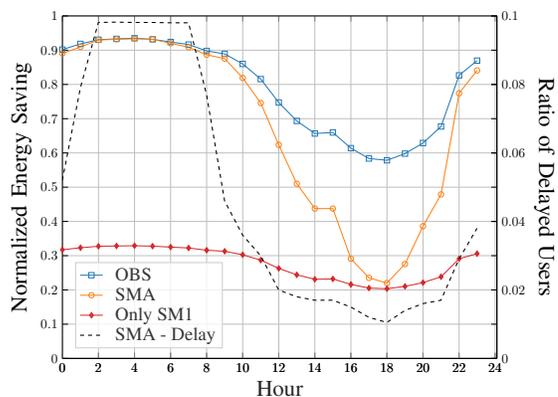
\begin{figure}[!t]
    \centering
    \resizebox{.85\columnwidth}{!}{
%
%

\definecolor{color0}{rgb}{0.12156862745098,0.466666666666667,0.705882352941177}
\definecolor{color1}{rgb}{1,0.498039215686275,0.0549019607843137}
\definecolor{color3}{rgb}{0.83921568627451,0.152941176470588,0.156862745098039}

\begin{tikzpicture}

\begin{axis}[%
width=4.521in,
height=3.566in,
at={(0.758in,0.481in)},
scale only axis,
xmin=-0, xmax=24.1,
xtick distance={2},
xlabel style={font=\color{white!15!black}},
xlabel={\LARGE Hour},
separate axis lines,
every outer y axis line/.append style={black},
every y tick label/.append style={font=\color{black}},
every y tick/.append style={black},
ymin=0,
ymax=1.0,
ylabel style={font=\color{black}},
ylabel={\LARGE Normalized Energy Saving},
axis background/.style={fill=white},
xmajorgrids,
ymajorgrids
]

\addplot [semithick, color0, mark=square, mark size=2, mark options={solid}]
table {%
0    0.9019
1    0.9184
2    0.9306
3    0.9331
4    0.9349
5    0.9317
6    0.9240
7    0.9165
8    0.8979
9    0.8892
10   0.8599
11   0.8156
12    0.7473
13    0.6934
14    0.6567
15    0.6597
16    0.6139
17    0.5837
18    0.5785
19    0.5980
20    0.6289
21    0.6772
22    0.8266
23    0.8695
}; \label{plotyyref:legen00}

\addplot [semithick, color1, mark=o, mark size=2, mark options={solid}]
table {%
0    0.8904
1    0.9096
2    0.9297
3    0.9318
4    0.9332
5    0.9317
6    0.9203
7    0.9085
8    0.8866
9    0.8755
10    0.8189
11    0.7457
12    0.6235
13    0.5101
14    0.4376
15    0.4374
16    0.2911
17    0.2352
18    0.2202
19    0.2756
20    0.3861
21    0.4790
22    0.7741
23    0.8410
};\label{plotyyref:legen01}

\addplot [semithick, color3, mark=diamond*, mark size=2, mark options={solid}]
table {%
0    0.3173
1    0.3231
2    0.3275
3    0.3283
4    0.3290
5    0.3278
6    0.3251
7    0.3225
8    0.3159
9    0.3129
10    0.3026
11    0.2870
12    0.2629
13    0.2440
14    0.2311
15    0.2321
16    0.2160
17    0.2054
18    0.2035
19    0.2104
20    0.2213
21    0.2383
22    0.2909
23    0.3060
};\label{plotyyref:legen02}
\end{axis}

\begin{axis}[%
width=4.521in,
height=3.566in,
at={(0.758in,0.481in)},
scale only axis,
xmin=-0,
xmax=24.1,
every outer y axis line/.append style={black},
every y tick label/.append style={font=\color{black}},
every y tick/.append style={black},
y tick label style={/pgf/number format/.cd,%
          scaled y ticks = false,
          fixed},
ymin=0,
ymax=0.1,
ytick distance={0.02},
ylabel style={font=\color{black}, rotate=180},
ylabel={\LARGE Ratio of Delayed Users},
axis x line*=bottom,
axis y line*=right,
legend pos=south west,
legend style={legend cell align=left, align=left,  fill opacity=0.8,
  draw opacity=1,
  text opacity=1,
  anchor=south west,
  draw=white!80!black }
]
\addlegendimage{/pgfplots/refstyle=plotyyref:legen00}
\addlegendentry{\Large OBS}

\addlegendimage{/pgfplots/refstyle=plotyyref:legen01}
\addlegendentry{\Large SMA}

\addlegendimage{/pgfplots/refstyle=plotyyref:legen02}
\addlegendentry{\Large Only SM1}

\addplot [thick,black, dashed]
table {%
0 0.052
1 0.079
2 0.0981
3 0.0982
4 0.0981
5 0.0981
6 0.0980
7 0.0980
8 0.077
9 0.046
10 0.036
11 0.03
12 0.020
13 0.018
14 0.017
15 0.017
16 0.015
17 0.012
18 0.0105
19 0.014
20 0.016
21 0.017
22 0.029
23 0.038
};\label{plotyyref:legen20}
\addlegendimage{/pgfplots/refstyle=plotyyref:legen20}
\addlegendentry{\Large SMA - Delay}
\end{axis}

\end{tikzpicture}%
}
    \caption{\small Daily profile of energy saving and ratio of delayed users.}
    \label{fig:24h}
\end{figure}

{\color{black}Fig. \ref{fig:24h} shows the  energy saving percentage of the SMA, only SM1 and OBS over a day. The values are calculated  with regards to the  energy consumption of the BS without using any SM and using the energy saving algorithms, i.e., OBS, Only SM1, and SMA.} Fig. \ref{fig:24h} illustrates that between 2:00-6:00 AM are the best hours of a day with the highest energy saving potentials.  The least opportunity for energy saving happens in the evening at about 18:00. It is worth mentioning the data set under consideration is for an area which is dominated by the industrial buildings. The traffic pattern and hence the energy saving patterns depend on the  type of area. For instance, in residential area, there might be a peak demand in the evening  when people are at home while  during the working hours network may experience low load within this area. 
During 2:00-6:00 AM, where the most energy saving is attained, SMA achieves considerable energy saving very close to the OBS. Slightly higher energy saving is achieved compared to OBS due to the delay tolerance allowing user arrivals when a BS is in deep sleep. OBS wakes up and does not allow any arrivals when BS is in sleep. During the peak hour, since there is a risk of delaying more number of users, SMA avoids going to deep sleeps and hence less energy saving is achieved compared to the OBS.  
 When only SM1 is activated, at very low load, the opportunity of saving energy is missed because we only use the shallowest SM. At very high load, there is less opportunity to activate longer and deeper SMs leading to similar performance between SMA and only-SM1 to avoid causing performance degradation. {\color{black} In this figure, we plot the ratio of delayed users compared to the total number of users in each hour. Between 2:00-6:00 AM, when the traffic is too low, the ratio of delayed users is higher because of two reasons, 1) the traffic load is low; therefore, even delaying one user increases the ratio significantly. 2) SMA achieves very high rewards due to high energy saving and choosing deeper SMs; hence, it accepts a small number of delaying users. Similar reasoning is applicable to other hours. It is worth noting that Fig. \ref{fig:24h} is very dependent on the value of $\alpha$. For instance, a small value of $\alpha$ scales up the ratio of delayed users in favor of saving more energy (please see Fig. 8). }

\subsubsection{Digital twin performance evaluation}
\begin{figure}[!t]
    \centering
    \resizebox{.43\textwidth}{!}{
%
%
\begin{tikzpicture}
\definecolor{color0}{rgb}{0.12156862745098,0.466666666666667,0.705882352941177}
\definecolor{color1}{rgb}{1,0.498039215686275,0.0549019607843137}
\pgfplotsset{
  log x ticks with fixed point/.style={
      xticklabel={
        \pgfkeys{/pgf/fpu=true}
        \pgfmathparse{exp(\tick)}%
        \pgfmathprintnumber[fixed relative, precision=3]{\pgfmathresult}
        \pgfkeys{/pgf/fpu=false}
      }
  }
}
\begin{axis}[%
legend cell align={left},
legend style={fill opacity=0.8, draw opacity=1, text opacity=1, draw=white!80!black},
width=4.521in,
height=3.566in,
at={(0.758in,0.481in)},
xmode=log,
    log ticks with fixed point,
    x filter/.code=\pgfmathparse{ #1},
domain=1:10,
xminorticks=true,
xlabel style={font=\color{white!15!black}},
xlabel={\large Arrival Rate},
ymin=0.1,
ymax=0.75,
ylabel style={font=\color{white!15!black}},
ylabel={\large Probability of Sleeping},
axis background/.style={fill=white},
xmajorgrids,
xminorgrids,
ymajorgrids
]
\addplot [color=color1, thick, mark=square, mark options={solid, color1}]
  table[row sep=crcr]{%
1	0.685009376953532\\
2	0.6432625\\
3	0.608160416666667\\
4	0.551106471816284\\
5	0.5069125\\
6	0.450477083333333\\
7	0.409981246092936\\
8	0.377154488517745\\
9	0.353755741127349\\
10	0.290960334029228\\
};
\addlegendentry{\Large SMA}
\addplot [color=color0, thick, mark = o]
  table[row sep=crcr]{%
1	0.6763740041681548\\
2	0.61761087396096\\
3	0.582255038273775\\
4	0.532734024509154\\
5	0.484867185814368\\
6	0.429680378084937\\
7	0.374874104745924\\
8	0.350927329892272\\
9	0.328411876330896\\
10	0.258273488675623\\
};
\addlegendentry{\Large DT model}
\end{axis}
\end{tikzpicture}%
}
    \caption{ \small Probability of sleeping in SMA vs estimated probability of sleeping via hidden Markov model. }
    \label{fig:markcomp}
\end{figure}
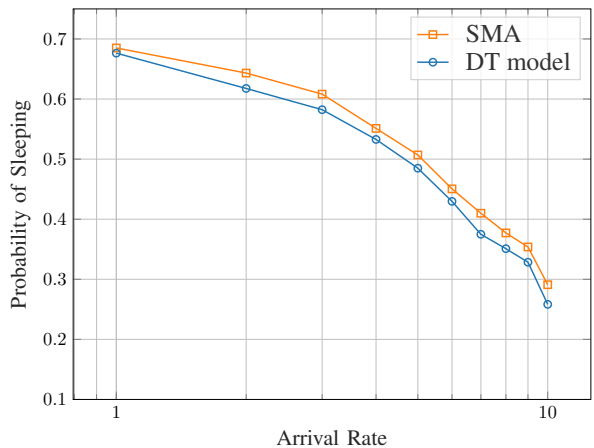

In Section \ref{sec:markov}, we present the DT model that can characterize the behavior of the sleep mode management algorithm. In order to calculate the probability of sleeping, we need to have information regarding the transition rates. This information can be obtained from the trained data, e.g., transition rates between sleep modes, and be updated in the physical network, e.g., arrival rate.  In Fig. \ref{fig:markcomp}, we compare the probability of sleeping derived in Equation \eqref{eq:prsleep} with the statistics collected from the implementation of the SMA. {\color{black} When running the SMA, for a given arrival rate, we measure the activity and inactivity time of the BSs. Assuming that during the inactivity time of the BS one of the SMs will be used\footnote{\color{black} We also assume that the random process of the user arrival is wide sense stationary.}, we calculate the probability of sleeping by dividing the inactivity duration by the total duration, i.e., activity plus inactivity time. It can be seen from this figure that  the DT  can properly model the behavior of the SMA for different arrival rates.}

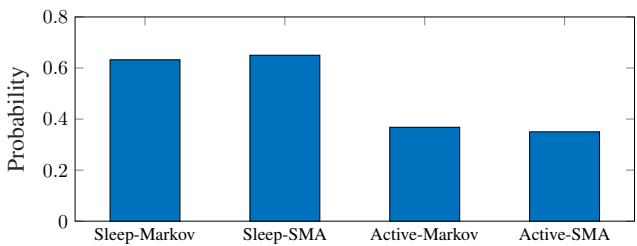
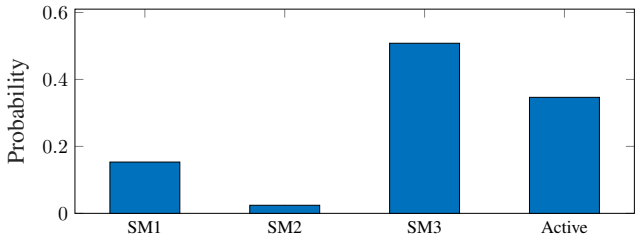
\begin{figure}[!t]
    \begin{subfigure}{.5\textwidth}
        \centering
        \hspace{-10mm}
        \scalebox{0.75}{
%
%
\definecolor{mycolor1}{rgb}{0.00000,0.44700,0.74100}%
\begin{tikzpicture}

\begin{axis}[%
width=2*1.952in,
height=0.4*3.566in,
at={(0.758in,0.481in)},
scale only axis,
bar width=0.5,
xmin=0.5,
xmax=4.5,
xtick={1,2,3,4},
xticklabels={{\small Sleep-Markov},{\small Sleep-SMA},{\small Active-Markov},{\small Active-SMA}},
ymin=0,
ymax=0.8,
ylabel style={font=\color{white!15!black}},
ylabel={\large Probability},
axis background/.style={fill=white}
]
\addplot[ybar stacked, fill=mycolor1, draw=black, area legend] table[row sep=crcr] {%
1	0.632255038273775\\
2	0.65\\
3	0.367886776936743\\
4	0.35\\
};

\end{axis}

\end{tikzpicture}%
}
        \caption{ \small Comparison of hidden Markov model estimation with SMA in terms of sleeping probability. }
    \label{fig:smMarDQN}
    \end{subfigure}\\
    \begin{subfigure}{.5\textwidth}
        \centering
        \hspace{-10mm}
        \scalebox{0.75}{\definecolor{mycolor1}{rgb}{0.00000,0.44700,0.74100}%
\begin{tikzpicture}
 
\begin{axis}[%
width=2*1.952in,
height=0.4*3.566in,
at={(3.327in,0.481in)},
scale only axis,
bar width=0.5,
xmin=0.5,
xmax=4.5,
xtick={1,2,3,4},
xticklabels={{\small SM1},{\small SM2},{\small SM3},{\small Active}},
ymin=0,
ymax=0.61,
ylabel style={font=\color{white!15!black}},
ylabel={\large Probability},
axis background/.style={fill=white}
]
\addplot[ybar stacked, fill=mycolor1, draw=black, area legend] table[row sep=crcr] {%
1	0.153234169220532\\
2	0.0241616443266086\\
3	0.508093393947166\\
4	0.34652607716211\\
};
\end{axis}
\end{tikzpicture}%
}
        \caption{\small  Breakdown of probability of sleeping for each sleep modes using DT. }
    \label{fig:MarDQN}
    \end{subfigure}
 \caption{\small Comparison of hidden Markov model estimation and SMA performance.}
\label{fig:MarDQNComp}
\end{figure}

 In Fig. \ref{fig:MarDQNComp}, we illustrate the breakdown of probability of being in each sleep mode. {\color{black} In Fig. \ref{fig:smMarDQN}, we compare the probability of sleeping and being active from the analysis, i.e., Markov model, and simulation, i.e., SMA. In Fig. \ref{fig:MarDQN}, utilizing the hidden Markov model, we illustrated the breakdown of the probability of being in each of the possible operating state of BSs, i.e., three sleep modes and being active. Assuming that the user arrival random process is wide sense stationary, the probability of being in each state corresponds to the fraction of the time BS is in each of the operating states. This information can be used to determine the most visited state by the BS for further energy saving improvement.   }

\subsubsection{RDM performance}
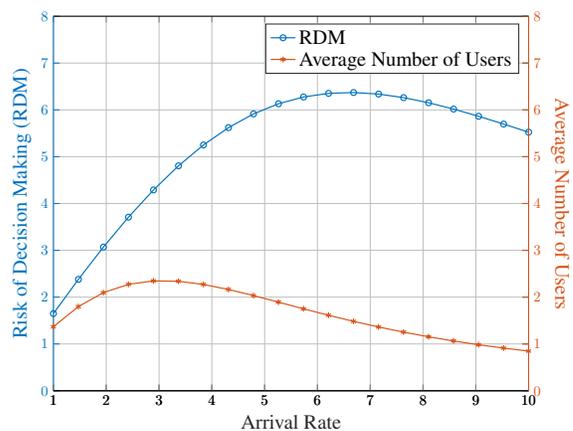
\begin{figure}[!t]
        \centering
        \resizebox{.85\columnwidth}{!}{
%
%
\definecolor{mycolor1}{rgb}{0.00000,0.44700,0.74100}%
\definecolor{mycolor2}{rgb}{0.85000,0.32500,0.09800}%
\begin{tikzpicture}

\begin{axis}[%
width=4.521in,
height=3.566in,
at={(0.758in,0.481in)},
scale only axis,
xmin=1,
xmax=10,
xlabel style={font=\color{white!15!black}},
xlabel={\Large Arrival Rate},
separate axis lines,
every outer y axis line/.append style={mycolor1},
every y tick label/.append style={font=\color{mycolor1}},
every y tick/.append style={mycolor1},
ymin=0,
ymax=8,
ytick={0, 1, 2, 3, 4, 5, 6, 7, 8},
ylabel style={font=\color{mycolor1}},
ylabel={\Large  Risk of Decision Making (RDM)},
axis background/.style={fill=white},
xmajorgrids,
ymajorgrids
]
\addplot [color=mycolor1,thick, mark=o, mark options={solid, mycolor1}, forget plot]
  table[row sep=crcr]{%
1	1.64954783730381\\
1.47368421052632	2.38045229672412\\
1.94736842105263	3.06920227509431\\
2.42105263157895	3.70811816543422\\
2.89473684210526	4.28936488694275\\
3.36842105263158	4.8056214527571\\
3.84210526315789	5.25081473913339\\
4.31578947368421	5.62079345010978\\
4.78947368421053	5.91382199870413\\
5.26315789473684	6.13080566550277\\
5.73684210526316	6.27521322509097\\
6.21052631578947	6.35272519199781\\
6.68421052631579	6.37068603226392\\
7.15789473684211	6.3374638804707\\
7.63157894736842	6.26181870332768\\
8.10526315789474	6.15235624317752\\
8.57894736842105	6.01711195390889\\
9.05263157894737	5.86327729810209\\
9.52631578947368	5.69705697135489\\
10	5.52363201143269\\
};
\label{plotyyref:leg1}

\end{axis}

\begin{axis}[%
width=4.521in,
height=3.566in,
at={(0.758in,0.481in)},
scale only axis,
xmin=1,
xmax=10,
every outer y axis line/.append style={mycolor2},
every y tick label/.append style={font=\color{mycolor2}},
every y tick/.append style={mycolor2},
ymin=0,
ymax=8,
ytick={0, 1, 2, 3, 4, 5, 6, 7, 8},
ylabel style={font=\color{mycolor2},rotate=180},
ylabel={\Large Average Number of Users},
axis x line*=bottom,
axis y line*=right,
legend style={legend cell align=left, align=left, draw=white!15!black}
]
\addlegendimage{/pgfplots/refstyle=plotyyref:leg1}
\addlegendentry{\Large  RDM}
\addplot [color=mycolor2, thick,mark=asterisk, mark options={solid, mycolor2}]
  table[row sep=crcr]{%
1	1.3668921480029\\
1.47368421052632	1.79877611048693\\
1.94736842105263	2.09664701851843\\
2.42105263157895	2.27335690878675\\
2.89473684210526	2.3470841630787\\
3.36842105263158	2.33924094171555\\
3.84210526315789	2.27172961280483\\
4.31578947368421	2.16453331166289\\
4.78947368421053	2.0342248913716\\
5.26315789473684	1.89346163116451\\
5.73684210526316	1.75120763060625\\
6.21052631578947	1.61334320329785\\
6.68421052631579	1.4833873034431\\
7.15789473684211	1.36316899900127\\
7.63157894736842	1.253376756779\\
8.10526315789474	1.15397251194659\\
8.57894736842105	1.06448601771074\\
9.05263157894737	0.984214551060834\\
9.52631578947368	0.912352860518281\\
10	0.84807413327782\\
};
\addlegendentry{\Large Average Number of Users}

\end{axis}

\end{tikzpicture}%
}
        \caption{\small Comparison of RDM and the average number of users experiencing the serving delay.}
    \label{fig:RDM1}
\end{figure}

According to the discussion in Section \ref{sec:RDMa}, the expected number of users cannot represent the risk of  decision making. In Fig. \ref{fig:RDM1}, we show that the RDM metric defined in Equation \eqref{eq:RDM} is capable of reflecting the risk of decision making in BS sleeping. When the arrival rate is high, i.e., BS is under the high load, BS is active for most of the times and less number of users should wait for BS to wake up and serve them. On the other hand, if the arrival rate is low, fewer number of users are active in the network and hence less number of users might wait until they are served. In both cases, the number of delayed users due to BS sleeping are small, however, the risk of BS sleeping is higher in the latter case. According to Fig. \ref{fig:RDM1}, unlike the average number of users metric, the RDM metric can perfectly differentiate between these two cases and show higher risk at the higher arrival rates.  
 

\subsubsection{Abnormal Traffic Behavior}

\begin{figure}[!t]
    \centering
    \resizebox{.85\columnwidth}{!}{\input{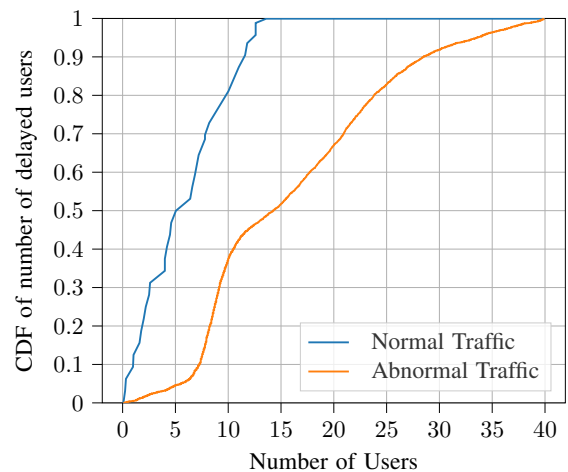}}
    \caption{\small CDF of delayed users}
    \label{fig:cdf}
\end{figure}

Despite the appealing performance of ML-based algorithms, their performance under abnormal situations, such as unseen abrupt changes, must be investigated. In Fig. \ref{fig:cdf}, we illustrate the cumulative distribution function (CDF) of the number of users who have been delayed due to BS sleeping in duration of one hour for two types of traffic, 1) normal traffic, and 2) abnormal traffic.  Normal traffic is a  traffic that the algorithm is trained over same data set. The abnormal traffic is an input traffic which is  generated from a different data set.   According to Fig. \ref{fig:cdf}, with $80$ $\%$ probability, less than $10$ requests are delayed in the observation period while in the abnormal traffic with probability of $50$ $\%$  more than $15$ users are delayed during the same observation period. According to this figure, with very high probability, large number of users are being delayed due to BS sleeping under abnormal traffic circumstances. Therefore, there is a need to detect the abnormal situations and avoid the risk of delaying large number of users.

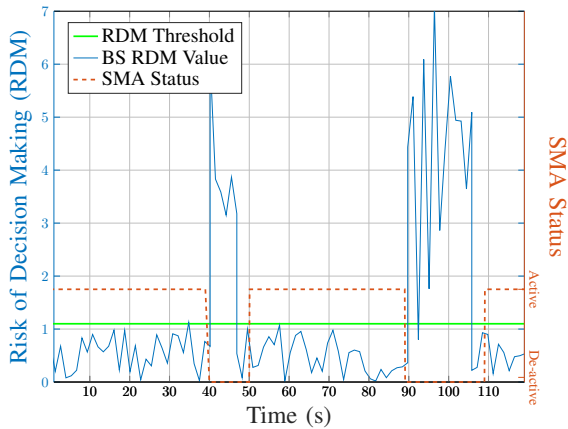
\begin{figure}[!t]
    \centering
    \resizebox{.85\columnwidth}{!}{
%
%
\definecolor{mycolor1}{rgb}{0.00000,0.44700,0.74100}%
\definecolor{mycolor2}{rgb}{0.85000,0.32500,0.09800}%
\begin{tikzpicture}

\begin{axis}[%
width=4.521in,
height=3.566in,
at={(0.758in,0.481in)},
scale only axis,
xmin=1,
xmax=119,
xlabel style={font=\color{white!15!black}},
xlabel={\LARGE Time (s)},
separate axis lines,
every outer y axis line/.append style={mycolor1},
every y tick label/.append style={font=\color{mycolor1}},
every y tick/.append style={mycolor1},
ymin=0,
ymax=7,
ytick={0, 1, 2, 3, 4, 5, 6, 7, 8},
ylabel style={font=\color{mycolor1}},
ylabel={\LARGE  Risk of Decision Making (RDM)},
axis background/.style={fill=white},
xmajorgrids,
ymajorgrids
]
\addplot [very thick, green]
table {%
0 1.1
120 1.1
};\label{plotyyref:leg00}
\addplot [ mycolor1] 
table {%
0 0.726047039031982
1.33928573131561 0.202803015708923
2.67857146263123 0.676543354988098
4.01785707473755 0.0794864892959595
5.35714292526245 0.118348002433777
6.69642877578735 0.221996545791626
8.0357141494751 0.831878423690796
9.375 0.566325902938843
10.7142858505249 0.897900104522705
12.0535717010498 0.664900779724121
13.3928575515747 0.570488452911377
14.7321424484253 0.675602674484253
16.0714282989502 0.978737115859985
17.4107151031494 0.226247549057007
18.75 0.975863933563232
20.0892848968506 0.196467518806458
21.4285717010498 0.684398174285889
22.7678565979004 0.0418508052825928
24.1071434020996 0.431078433990479
25.4464282989502 0.305169939994812
26.7857151031494 0.89141058921814
28.125 0.642678737640381
29.4642848968506 0.355978727340698
30.8035717010498 0.911053061485291
32.1428565979004 0.872771024703979
33.4821434020996 0.556448221206665
34.8214302062988 1.124150907993317
36.1607131958008 0.346898555755615
37.5 0.0252026319503784
38.8392868041992 0.764960408210754
40.1785697937012 0.673962831497192
40.1785697937012 6.09933471679688
41.5178565979004 3.83076906204224
42.8571434020996 3.58711004257202
44.1964302062988 3.15038681030273
45.5357131958008 3.86774945259094
46.875 3.17807412147522
46.875 0.537414789199829
48.2142868041992 0.0808868408203125
49.5535697937012 0.998383283615112
50.8928565979004 0.276416540145874
52.2321434020996 0.312986135482788
53.5714302062988 0.659948825836182
54.9107131958008 0.856030702590942
56.25 0.706021785736084
57.5892868041992 1.07463324069977
58.9285697937012 0.0266883373260498
60.2678565979004 0.555846214294434
61.6071434020996 0.880176067352295
62.9464302062988 0.954963684082031
64.2857131958008 0.612878084182739
65.625 0.177628636360168
66.9642868041992 0.453728318214417
68.3035736083984 0.203011035919189
69.6428604125977 0.736898899078369
70.9821395874023 0.976723194122314
72.3214263916016 0.583666563034058
73.6607131958008 0.0452547073364258
75 0.555450916290283
76.3392868041992 0.603331685066223
77.6785736083984 0.573593854904175
79.0178604125977 0.20914888381958
80.3571395874023 0.0618141889572144
81.6964263916016 0.0181591510772705
83.0357131958008 0.237341642379761
84.375 0.0822173357009888
85.7142868041992 0.212102293968201
87.0535736083984 0.2700514793396
88.3928604125977 0.286688685417175
89.7321395874023 0.364221096038818
89.7321395874023 4.43500471115112
91.0714263916016 5.38832426071167
92.4107131958008 0.80162787437439
93.75 6.09138488769531
95.0892868041992 1.75977861881256
96.4285736083984 7.11339807510376
97.7678604125977 2.86094737052917
99.1071395874023 4.4129204750061
100.446426391602 5.77471542358398
101.785713195801 4.94131278991699
103.125 4.92601633071899
104.464286804199 3.64394521713257
105.803573608398 5.09261560440063
105.803573608398 0.221128940582275
107.142860412598 0.277041912078857
108.482139587402 0.930987000465393
109.821426391602 0.898050785064697
111.160713195801 0.161330938339233
112.5 0.713524341583252
113.839286804199 0.556922793388367
115.178573608398 0.215091705322266
116.517860412598 0.47816002368927
117.857139587402 0.496978998184204
119.196426391602 0.537155866622925
120.535713195801 0.273958802223206
};\label{plotyyref:leg20}
\end{axis}

\begin{axis}[%
width=4.521in,
height=3.566in,
at={(0.758in,0.481in)},
scale only axis,
xmin=1,
xmax=119,
every outer y axis line/.append style={mycolor2},
every y tick label/.append style={font=\color{mycolor2}},
every y tick/.append style={mycolor2},
ymin=0,
ymax=8,
ytick={0.1, 2},
yticklabels={{De-active},{Active}, {},{},{}, {}, {}, {}},
ylabel style={font=\color{mycolor2}, rotate=180},
yticklabel style ={rotate = 270},
ylabel={\LARGE SMA Status},
axis x line*=bottom,
axis y line*=right,
legend pos=north west,
legend style={legend cell align=left, align=left, draw=white!15!black}
]
\addlegendimage{/pgfplots/refstyle=plotyyref:leg00}
\addlegendentry{\Large  RDM Threshold}
\addlegendimage{/pgfplots/refstyle=plotyyref:leg20},
\addlegendentry{\Large BS RDM Value}

\addplot [very thick, dashed,mycolor2]
table {%
0 2
39 2
40 0
49 0 
50 0
50.1 2
88 2
89 2
89.1 0
107 0
109 0
109.1 2
120 2
};\label{plotyyref:leg22}
\addlegendimage{/pgfplots/refstyle=plotyyref:leg22}
\addlegendentry{\Large SMA Status}

\end{axis}

\end{tikzpicture}%
}
    \caption{\small  Performance of risk management algorithm}
   \label{fig:RDM_dec}
\end{figure}

 In Fig. \ref{fig:RDM_dec}, we plot the performance of Algorithm \ref{alg:RDM}. Using the procedure in Algorithm \ref{alg:RDM}, BS calculates the risk value and when the risk is high it temporarily deactivates SMs. When SMs are deactivated, BS calculates the potential risks and when the risk is small again for a while, i.e., for duration of $T$, it activates the SMs again.  
 {\color{black} The performance of the risk management algorithm is dependent on the RDM threshold which triggers the (de/re)activation of the SMA. When RDM threshold is high, BS accepts more risk and hence more number of users can be delayed. Hence SMA will be active for a longer duration. On the other hand, very low value of RDM threshold makes the BS very sensitive to delaying the users and hence make less use of SMA to save energy. This threshold can be set by the operator depending on the time and location of the BS. }
 
    


\vspace{-4pt}
\section{Conclusion} \label{sec:con}
In this paper, we proposed a deep Q-learning algorithm to reduce the energy consumption of 5G base stations using  multi-level sleep modes (SMs). In particular, we considered 3 SMs each with distinct (de)activation time and energy consumption. Since there is always a risk associated to machine learning algorithms, we proposed a framework to predict and manage the associated risk. First, we defined a novel metric to quantify the risk of decision making (RDM).  We demonstrated that the proposed metric can represent the risk of taking non-optimal decisions. Then, we  proposed a digital twin model (DT)  that can predict the performance of the proposed DQN algorithm by estimating the expected probability of sleeping and the expected RDM. Finally, thanks to the proposed DT, BS can detect the abnormal behavior of traffic data and deactivate the SMs  to avoid any extra performance degradation.  We evaluated the performance of proposed algorithm using real network data obtained from a BS in Stockholm and  compared it with the baselines. Simulation results confirmed that with the proposed DT, we can estimate the performance of the proposed SM management algorithm. With the help of this model, we can avoid incurring large delays to the users due to abnormal behavior of input traffic. Moreover, the simulation results showed that considerable energy saving can be achieved with a good compromise with the serving delay, considering that number of users that will be delayed can be controlled thanks to the DT assisted learning mechanism.
\appendix
\section{Balance Equations for the Proposed Markov Model}\label{sec:appA}
In this section, we write the balance equation for the Markov process depicted in Fig. \ref{fig:markovChain}. 
\begin{align} \label{eq:balance_s}
 &\nu_{1,2}(\tau \!\!+\!\! \pi_{1,3} \!\!+\!\! \pi_{1,2}) \!=\!\nu_{2,2} \pi_{2,1} \!\!+\!\! \nu_{3,2} \pi_{3,1} \!\!+\!\! \zeta \nu_{1,1} \!\!+\!\! u_{1,2} \mu_{1,2}\\
 &\nu_{2,2}(\tau\!\!+\!\! \pi_{2,3} \!\!+\!\! \pi_{2,1}) \!=\! \nu_{1,2} \pi_{1,2} \!\!+\!\! \nu_{3,2} \pi_{3,2} \!\!+\!\! \zeta \nu_{2,1} \!\!+\!\! u_{1,2} \mu_{2,2}\\ 
 &\nu_{3,2}(\tau \!\!+\!\! \pi_{3,1} \!\!+\!\! \pi_{3,2}) \!=\! \nu_{1,2} \pi_{1,3} \!\!+\!\! \nu_{2,2} \pi_{2,3}\!\! +\!\! \zeta \nu_{3,1} \!\!+\!\! u_{1,2} \mu_{3,2}\\
 &\nu_{1,1}(\lambda_1 + \zeta) = u_{1,1}\mu_{1,1} + \tau \nu_{1,2}\\
 &\nu_{2,1}(\lambda_2 + \zeta) = u_{1,1}\mu_{2,1} + \tau u_{2,2}\\
 &\nu_{3,1}(\lambda_3 + \zeta) = u_{1,1}\mu_{3,1} + \tau u_{3,2}\\
 &u_{1,1}\!\left(\zeta\!\!+\!\!\mu_{1,1}\!\!+\!\!\mu_{2,1}\!\!+\!\!\mu_{3,1}\!\!+\!\!\lambda\right)\!\!=\!\! \tau u_{1,2} \!\!+\!\! \lambda\!\left(\nu_{1,1}\!\!+\!\!  \nu_{2,1}\!\!+\!\! \nu_{3,1}\right) \!\!+\!\! \mu u_{2,1}\\
 &u_{1,2}(\tau\!+\!\mu_{1,2}\!+\!\mu_{2,2}\!+\!\mu_{3,2})=\zeta u_{1,1} \!+\! \mu u_{2,2}\\
 &(\zeta \!\!+\!\! \mu\!\!+\!\!\lambda) u_{m,1} \!\!=\!\! \lambda u_{m\!-\!1,1} \!\!+\!\! \mu u_{m\!+\!1,1} \!\!+\!\! \tau u_{m,2},\ \! m \!\in\! \left[2, M\!\!-\!\!1\right] \\
 &(\tau+\mu)u_{m,2} = \mu u_{m+1,2}+\zeta u_{m,1}, \qquad\ \ \ \hspace{0.1mm}  \!m\! \in\! \left[2, M\!\!-\!\!1\right] \\
 &(\mu+\zeta) u_{M,1}=\lambda u_{M-1,1} + \tau u_{M,2} \\
 &(\tau + \mu)u_{M,2}=\zeta u_{M,1}\\
  &\sum_{j=1}^{2}\left(\sum_{i=1}^{3}\nu_{i,j} + \sum_{m=1}^{M} u_{m,j}\right) =1 \label{eq:balance_end}
\end{align}
In Equations \eqref{eq:balance_s}-\eqref{eq:balance_end}, we have $2M+6$ unknowns and $2M+6$ independent equations. Since all equations are linear, well-known techniques can solve this system of equations. Assume matrix A be the coefficient matrix, X be the vector of unknowns, and B be the vector of constants. One can easily solve the linear equation sets, i.e., $A X = B$, using existing solvers, such as Matlab or IBM CPLEX. 


\bibliographystyle{IEEEtran}
\bibliography{IEEEabrv,Ref}

\end{document}